\documentclass[]{jfm}

\usepackage{amsmath,mathtools}
\usepackage{graphicx}
\usepackage{wrapfig}
\usepackage{lscape}
\usepackage{rotating}
\usepackage{newtxtext}
\usepackage{newtxmath}
\usepackage{natbib}
\usepackage{hyperref}
\usepackage{float}
\usepackage{caption}
\usepackage{subcaption}
\usepackage{sidecap}
\usepackage{soul}
\usepackage{tabularx}
\usepackage[graphicx]{realboxes}
\usepackage{sidecap}
\hypersetup{
    colorlinks = true,
    urlcolor   = blue,
    citecolor  = black,
}

\newcommand{\RomanNumeralCaps}[1]
\linenumbers
\title{The dynamics of fibers dispersed in viscoelastic turbulent flows}

\author{M. S. Aswathy,
  Marco Edoardo Rosti\corresp{\email{marco.rosti@oist.jp}}}
\affiliation{Complex Fluids and Flows Unit, Okinawa Institute of Science and Technology Graduate University, 1919-1 Tancha, Onna-son, Okinawa 904-0495, Japan}
\begin{document}

\maketitle

\begin{abstract}

\end{abstract}

\begin{keywords}
% Authors should not enter keywords on the manuscript, as these must be chosen by the author during the online submission process and will then be added during the typesetting process (see \href{https://www.cambridge.org/core/journals/journal-of-fluid-mechanics/information/list-of-keywords}{Keyword PDF} for the full list).  Other classifications will be added at the same time.
\end{keywords}

%{\bf MSC Codes }  {\it(Optional)} Please enter your MSC Codes here
\begin{abstract}
This study explores the dynamics of finite-size fibers suspended freely in a viscoelastic turbulent flow. For a fiber suspended in Newtonian flows, two different flapping regimes were identified by \cite{rosti2018flexible}: one dominated by time scales from the flow, and another by time scales associated with its natural frequency. \textcolor{black}{We explore in this work, how the fiber dynamics is modified by the elasticity of the carrier fluid. For this, we perform Direct Numerical Simulations (DNS) of a two-way coupled fiber-fluid system in a parametric space spanning different Deborah numbers, fiber bending stiffness (flexible to rigid) and linear density difference between the fiber and the flow (neutrally-bouyant to denser-than-fluid fibers). We examine how these parameters influence various fiber characteristics such as the frequency of flapping, curvature, and alignment with the fluid strain and polymer stretching directions.} Results reveal that the neutrally-bouyant fibers, depending on their flexibility, oscillate with large and small time scales transpiring from the flow, but the smaller time-scales are suppressed as the polymer elasticity increases. Polymer stretching is uncommunicative to denser-than-fluid fibers, which flap with large time scales from the flow when flexible and with their natural frequency when rigid. Thus, the characteristic elastic time scale has a subdominant effect when the fibers are neutrally-bouyant, while its effect is absent when the fibers become more inertial. In addition, we also explore the fiber's bending curvature and its preferential alignment with the flow to identify the other roles of viscoelasticity in modifying the coupled fluid-structure dynamics. Inertial fibers have larger curvatures and are less responsive to the polymer presence, whereas the neutrally-bouyant fibers show quantitative changes. These perceptible passivity of the denser fibers are again reflected in the way they preferentially align with the polymeric stretching directions: the neutrally-bouyant fibers show a higher alignment with the polymer stretching directions compared to the denser ones. In a nutshell, the polymers exert a larger influence on neutrally-bouyant fibers, that are more reflective of the polymeric influence in the flow. The study comprehensively addresses the interplay between polymer elasticity and the fiber structural properties in determining its response behaviour in an elasto-inertial turbulent flow.
\end{abstract}

\section{Introduction}
\label{sec:headings}

Systems involving filament-like structures interacting with fluids are common in nature and many industrial processes, such as microplastics in aquatic environments and pulp production in paper-making  \citep{guasto2012fluid,lundell2011fluid,du2019dynamics,carichino2021computational}, and are studied also due to their similarities with the dynamics of complex systems, such as swimming fish, flapping flags \citep{zhang2000flexible,tian2013role}, etc. Studies involving Newtonian fluids \textcolor{black}{\citep {brouzet2014flexible,allende2018stretching,parsa2012rotation,ni2015measurements,kuperman2019inertial,sulaiman2019numerical,zuk2021universal}} are more common compared to the research done on filaments interacting with non-Newtonian viscoelastic fluids. However, filament-fluid interactions in the background of viscoelasticity or `polymeric' influence have gathered more attention recently owing to their presence in many biological and industrial scenarios; to name a few, fluid transport in biological and technological scenarios involving confined environments, such as cilia which transport trapped particles out of the lungs from a viscoelastic mucus layer \citep{guo2017computational}, filament-like biological polymers such as actin \citep{gisler1999scaling}, pulp fiber suspensions in paper-making industry \citep{hearle2008physical} and in the development of nanocomposite materials wherein the nanotubes \citep{hobbie2003orientation}
  %has a good picture for slide.chek the paper.
are essentially microscopic fibers, \textcolor{black}{suspensions of which can also induce flow-induced gelation and shear thickening \citep{perazzo2017flow}}.

In this work, we study the dynamics of elongated finite-size fibers immersed in a tri-periodic domain forced by a cellular flow, thus dealing with a fiber-fluid-viscoelastic system in a highly turbulent flow regime. \textcolor{black}{By `finite size', we mean that the fiber has a finite length, comparable to length scales in the inertial range of turbulence.} The presence of polymers introduces also (at least) an additional dimensionless number, called the Deborah number $De$, which is the ratio of the polymeric relaxation time scale $\Lambda$ over a characteristic time scale of the flow, say $L_{0}/U_{rms_{0}}$, $L_{0}$ being the integral length scale of the flow and $U_{rms_{0}}$ the root mean square flow velocity. \cite{yang2017dynamics} modeled the motion of a single fiber dispersed in a polymeric cellular flow using two-dimensional computations at very low Reynolds number and noticed that the fiber in a Newtonian fluid travels faster and buckles earlier in comparison to their viscoelastic counterparts. %The polymer stresses were concentrated near the end when the fiber was straight whereas it got distributed along its length when it buckled.
Often experiments have been carried out to study the dynamics of fibers in polymer-laden flows subject to simple shear flows, where there are interests in probing whether the fiber gets aligned to the vorticity or the flow directions~\citep{iso1996orientation2,iso1996orientation}; there have been definite industrial interests in knowing how the shear flow orients the semiflexible nanotubes and how the elasticity of the viscoelastic melt influences the latter \citep{hobbie2003orientation}. Viscoelastic flows interacting with non-massless deforming structures were studied with an IB-LBM method for the first time by \cite{ma2020immersed}, who found that viscoelasticity can hinder the three-dimensional flapping motion of flags. Viscoelasticity is also reported to alter the beating patterns of swimmers (filament-like), which in turn influences their swimming velocities and the power dissipated \citep{fu2008beating}. 
 
The above studies irrefutably show that the complexities associated with viscoelasticity or non-Newtonian effects are intrinsic to most fiber-fluid interaction systems. However, most of the literature has attempted to test the fiber dynamics under very low Reynolds number flows and two-dimensional flow conditions, which although can provide clues to the key dynamics, do not unravel the complications arising out of fluid turbulence. In this context, we attempt to track the fiber dynamics for the first time using three-dimensional direct numerical simulations (DNS) in a homogeneous isotropic turbulent viscoelastic flow, to investigate how the polymeric fluid turbulence influences the fiber dynamics and to analyze if there are qualitative features associated with this system which are not captured by the two-dimensional simplifications in previous studies. Indeed, filament-fluid interactions in past studies conducted in Newtonian turbulent flows have shown that fiber bending stiffness and the linear density difference between the fiber and flow play imperative roles in deciding the fiber's flapping frequency. \textcolor{black}{\cite{oehmke2021spinning} measured Lagrangian time scales (related to spinning and tumbling) of inertial fibers in turbulence and reported that they scale with length and diameter of the fiber. It was observed by \cite{rosti2018flexible,olivieri2022fully}} that under the right parametric combinations, the flexible fibers flapped approximately with the turbulent eddy frequency at its length scale, and the stiffer ones flapped with their inherent natural frequency. \textcolor{black}{This observation suggested that fibers could be used to measure the two point statistics of the flow, at least in certain parametric regimes.} It is also in the interest of the present work to understand if viscoelasticity can modify \textcolor{black}{the above discussed} known dynamics.

We attempt, for the first time, a systematic approach to study the fully coupled \textcolor{black}{fiber-fluid system at a high Reynolds number turbulent flow, where the fluid is viscoelastic.  %(the Taylor microscale Reynolds number $Re_{\lambda} \approx 310$) 
The simulations are performed for a homogeneous isotropic flow (HIT)} subject to a cellular forcing, with parametric variations in the polymer relaxation time, the fiber's bending stiffness (flexible and rigid fibers) and the linear density difference between fiber and flow (almost neutrally bouyant and denser-than-fluid fibers). The fiber dynamics is tracked through its flapping frequency, the bending curvature, and its alignment with the relevant fluid quantities to learn if viscoelasticity influences its course of action. % remember to write abt sections here.
\textcolor{black}{The main open question tackled by our study is the following:
\begin{itemize}
\item What is the dynamical response of a fiber (with rest length in the turbulent inertial range of scales) when it is dispersed in a turbulent viscoelastic flow? 
\newline
More specifically, we would like to focus on the fiber deformation measured through its end-end displacement and look at various aspects of it, such as:
\item How does the temporal and frequency dynamics of suspended fibers get influenced in a parametric plane comprising of its flexibility, its linear density difference with respect to the fluid, and most importantly, with the polymer relaxation time?
\item Can one quantify the effects of viscoleasticity on the fiber deformation (e.g. through its curvature and bending energy)?
\item  Is the alignment of the fiber with the flow (e.g. the principal directions of the strain rate tensor) altered by the presence of polymers?
\end{itemize}}
 Section 2 describes the methodology and details of computations carried out to execute the study, 
Section 3 discusses the results and Section 4 concludes the study.

\section {Methodology}
 %\begin{figure}
 \begin{figure}
   \centering \includegraphics[width=1\textwidth,trim={0cm 0 0 0},clip]{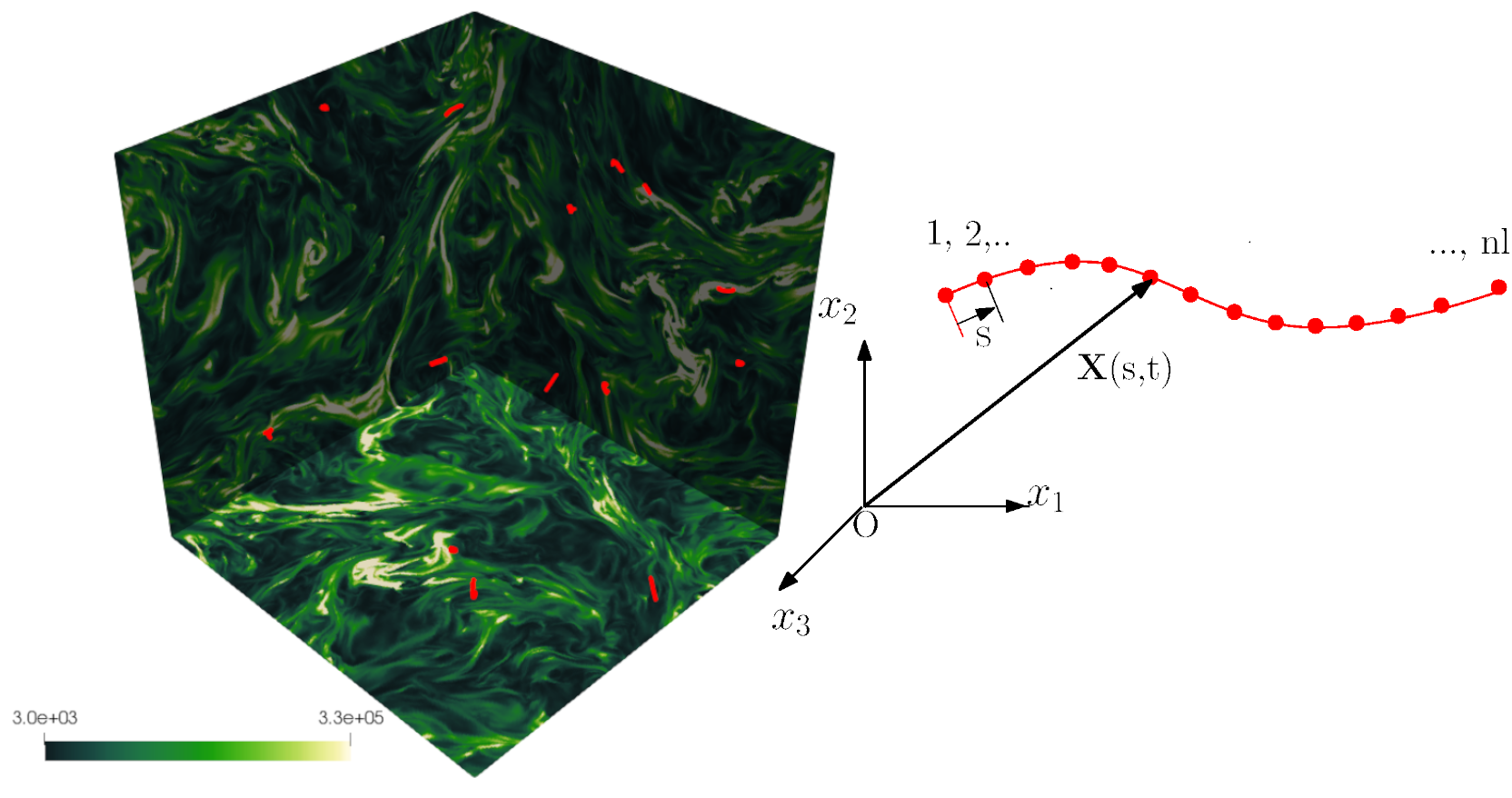}
    
% \end{figure}
%      \begin{subfigure}{0.4\textwidth}
%      %\vspace{-80cm}
% \includegraphics[width=1\textwidth,trim={0cm 0 0 0},clip] {figuresredo/s3.png}     
%      \end{subfigure}
          \caption{(a) A qualitative snapshot from the DNS simulations for $Re_{\lambda_{0}}$ $\approx$ 310 and Deborah number $De \approx$  7, where fibers of various rigidities are dispersed in a tri-periodic domain. The three backplanes are colored based on the trace of the polymer conformation tensor and the red lines represent the fibers. (b) Schematic of the fiber and of the Lagrangian points. }
           \label{fig:domain}  
 \end{figure}
We consider finite-size flexible fibers with various bending rigidities $\gamma$ and with lengths $c$ within the inertial range of scales, dispersed in a homogeneous isotropic turbulent flow in a cubic domain of size $L_{d}$ with periodic boundary conditions applied in all directions, as shown in Fig~\ref{fig:domain}a. The reference configuration is chosen such that the Taylor microscale Reynolds number of the corresponding single phase flow $\operatorname{Re}_{\lambda_{0}} \equiv U_{\mathrm{rms_{0}}} \lambda_{0} / \nu \approx 310$, where $U_{\mathrm{rms}}$ is the root-mean-square velocity, $\lambda_{0}$ is the Taylor micro-scale, and $\nu$ is the kinematic viscosity. %, with the Deborah number value being 7.
In the flow we inject a series of fibers with different flexibility, achieving a volume fraction $\Phi_{V} = V_{s}/V_{f} = 1.89 \times 10^{-5}$, defined as the ratio of the volume of the dispersed phase $V_{s} = Nc\pi d^2/4$ and that of the fluid phase $V_{f} = L_{d}^3$, where $N$ is the number of fibers, $d$ the diameter, and $c$ the unbent fiber length. % in a domain of length $L_{d}$ = $2 \pi$. %, as well as the mass fraction $M= m_s/(m_s +m_f)$, where $m_s = \rho_s V_s$ and $m_f = \rho_f V_f$ with $\rho_f$ being fluid density is the fluid mass,
This implies that the suspension is very dilute and hence the overall back-reaction effects on the flow from the fiber are minimal.
%, which has been already extensively used by the Authors for both flexible and rigid fibers, as well as dilute vs non-dilute fiber suspensions  where the mutual interaction between the solid and fluid phase is enforced indirectly by means of the singular force distributions $\mathbf{F}$ and $\mathbf{f}_{\text {fib }}$ acting on the fiber and flow, respectively 12 . 

Each fiber is modeled as a homogeneous, inextensible elastic filament evolving according to the Euler-Bernoulli beam equation in a Lagrangian framework as:
\begin{equation}
{\rho_{l}} \ddot{\mathbf{X}}=\partial_{s}\left(T \partial_{s}(\mathbf{X})\right)-\gamma \partial_{s}^{4} \mathbf{X}-\mathbf{F}, \\
\end{equation}
\begin{equation}
\partial_{s} \mathbf{X} \cdot \partial_{s} \mathbf{X} =1,
\end{equation}
where $\mathbf{X}(s, t)$ is the fiber position based on the curvilinear coordinate $s$ at time $t$, ${\rho_{l}} = \widetilde{\rho}_{\mathrm{s}} -\widetilde{\rho}_{\mathrm{f}}$ is the difference between the linear density of the solid and the fluid, $\gamma$ is the bending stiffness (defined as the product of the elastic modulus and the second moment of the area), $T$ is the tension enforcing the inextensibility, and $\boldsymbol{F}$ is the effect of the fluid acting on the fiber. Freely-moving boundary conditions are imposed at both fiber ends: $\left.\partial_{s s} \mathbf{X}\right|_{s=0, c}=\left.\partial_{s s s} \mathbf{X}\right|_{s=0, c}=\left.T\right|_{s=0, c}=0$. \textcolor{black}{In this work we consider both heavy and almost neutrally-buoyant fibers, with $\rho_{l}=1$ and $10^{-3}$, and we vary the bending stiffness over several order of magnitudes in the range $[10^{-8},10^{-2}]$. We can define a non-dimensional bending stiffness $\gamma^{\prime}$ by comparing the inertial and bending terms, i.e., $\gamma^{\prime}$ = $\frac{\gamma}{\rho_{l} U_{rms_{0}}^2 L_{0}^2}$ where $L_{0}$ is the integral length scale of the flow, and obtain approximately $\gamma^ {\prime} \in [10^{-8},10^{-1}]$ for neutrally-bouyant fibers and $\in [10^{-11},10^{-4}]$ for denser ones. These values are comparable to those achieved experimentally by \cite{gay2018characterisation} where $\gamma ^{\prime} \in [10^{-6}, 10^{6}]$.} \textcolor{black}{The schematic of the fiber is shown in Fig~\ref{fig:domain}b, where $x_{1}$, $x_{2}$, and $x_{3}$ are the coordinate axes corresponding to the Eulerian frame of reference; $nl$ is the number of lagrangian points on the fiber taken as 25 here. Suppose $\mathbf{X_{1}}$ and $\mathbf{X_{2}}$ are the Lagrangian coordinates of the fiber tips, the end-end displacement is defined as $\lvert \mathbf {X_{2}}- \mathbf {X_{1}} \rvert$. In this study, we characterise this quantity which represents an effective fiber deformation (mainly quantifying its bending). } % and $\mathbf{F}_{\text {col }}$ is a fiber-to-fiber collision forcing term. 

The fluid flow is governed by the incompressible Navier-Stokes equations for a viscoelastic fluid:
\begin{equation}
 \rho\left(\frac{\partial \boldsymbol{u}}{\partial t}+\boldsymbol{u} \cdot \nabla \boldsymbol{u}\right)=-\nabla p+\nabla \cdot\boldsymbol{\tau}+\rho({\boldsymbol{f}_{fib} +\boldsymbol{f}_{turb}}),
\end{equation}
\begin{equation}
 \nabla \cdot \boldsymbol{u}=0,
\end{equation}
where $\boldsymbol{u}(\mathbf{x}, t)$ and $p(\mathbf{x}, t)$ are the velocity and pressure fields, $\rho$ is the fluid density, $\boldsymbol{\tau}$ is the stress tensor, and $\mathbf{f}_{fib}$ is the feedback forcing from the fiber. $\mathbf{f}_{\mathrm{turb}}$ is the external forcing term used to sustain turbulence, chosen to be the Arnold-Beltrami-Childress (ABC) forcing \citep{dombre1986chaotic} given by  $\mathbf{f}_{\mathrm{turb}}=\nu(A \sin z+C \cos y, B \sin z+A \cos z, C \sin y+B \cos z)$ with \textcolor{black}{constant parameters $A$ = $B$ = $C$ =1}.

The stress tensor $\boldsymbol{\tau}$ is the sum of two contributions coming from the solvent and polymer \citep{comminal2015robust}, i.e.~$\boldsymbol{\tau}=\boldsymbol{\tau}_{\mathrm{S}}+\boldsymbol{\tau}_{\mathrm{P}}$, with $\boldsymbol{\tau}_{\mathrm{S}} =2 \eta_{\mathrm{s}} \boldsymbol{D}$ and $\boldsymbol{\tau}_{\mathrm{P}}=G_0 f_{\mathrm{S}}(\boldsymbol{c})$, where $\eta_{\mathrm{S}}=\beta \eta_t$ is the solvent viscosity, $\beta $ is the solvent to total viscosity ratio chosen to be equal to 0.9, and the rate of deformation tensor is $\boldsymbol{D}=\left(\nabla \boldsymbol{u}+\nabla \boldsymbol{u}^{\mathrm{T}}\right) / 2$. $G_0=(1-\beta) \eta_0 / \Lambda$ is the \textcolor{black}{polymeric} elastic modulus, $\Lambda$ the relaxation time and %material parameters of linear viscoelasticity; 
$f_S(\boldsymbol{c})$ is a strain function expressed in terms of the conformation tensor $\boldsymbol{c}$, which is representative of the orientation of the polymer chains. A matrix log-conformation formulation \citep{fattal2004constitutive,fattal2005time,izbassarov2018computational} is used to solve the above equation, where the variable $\boldsymbol{\Psi}$ = $\log  \boldsymbol{c}$ is invoked and the transport equation is modified as
\begin{equation}
\frac{\partial \boldsymbol{\Psi}}{\partial \boldsymbol{t}}+(\boldsymbol{u} \cdot \nabla) \boldsymbol{\Psi}-(\boldsymbol{\Omega} \boldsymbol{\Psi}-\boldsymbol{\Psi} \boldsymbol{\Omega})-2 \boldsymbol{E} \\
=-\frac{1}{\Lambda} \exp (-\boldsymbol{\Psi}) f_{\mathrm{R}}[\exp (\boldsymbol{\Psi})],
\end{equation}
where the matrices $\boldsymbol{\Omega}$ and $\boldsymbol{E}$ are pure rotation (anti-symmetric) and pure extension (symmetric traceless) matrix, obtained from the projection of the velocity gradient into the principal base of the stress tensor \citep{fattal2004constitutive}; $f_{R}$ is the relaxation function. In the present work, we consider Oldroyd-B fluids, which have the following strain and relaxation functions: $f_{\mathrm{S}}(\boldsymbol{c})=f_{\mathrm{R}}(\boldsymbol{c})=\boldsymbol{c}-\boldsymbol{I}$, where $\boldsymbol{I}$ is the identity matrix \citep{comminal2015robust}. The conformation tensor $\mathbf{c}$ is normalized such that at equilibrium, it turns out to be an identity matrix.% The Oldroyd-B model corresponds to the limiting case of linear-viscoelasticity, for most of the non-linear models. It is known for making unrealistic predictions in pure extensional deformations, due to the dissipative constitutive instability. 
% where $\eta_{\mathrm{S}}=\beta \eta_0$ is the solvent viscosity, $G_0=(1-\beta) \eta_0 / \lambda$ is the elastic moduli,
% $$
% \boldsymbol{D}=\left(\nabla \boldsymbol{u}+\nabla \boldsymbol{u}^{\mathrm{T}}\right) / 2
% $$
% $$
% \tau=\tau_{\mathrm{S}}+\tau_{\mathrm{P}},
% $$huang
% with
% $$
% \begin{aligned}
% &\tau_{\mathrm{S}}=2 \eta_{\mathrm{s}} \boldsymbol{D}, \\
% &\tau_{\mathrm{P}}=G_0 f_{\mathrm{S}}(\boldsymbol{c}),
% \end{aligned}
% $$
% where $\eta_{\mathrm{S}}=\beta \eta_0$ is the solvent viscosity, $G_0=(1-\beta) \eta_0 / \lambda$ is the elastic moduli,
% $$
% \boldsymbol{D}=\left(\nabla \boldsymbol{u}+\nabla \boldsymbol{u}^{\mathrm{T}}\right) / 2
% $$
% is the rate of deformation tensor, and $f_S(\boldsymbol{c})$ is a strain function expressed in terms of the conformation tensor $\boldsymbol{c}$. The steady-state viscosity $\eta_0$, the relaxation time $\lambda$, and the retardation ratio $\beta \in[0,1]$ are the material parameters of linear viscoelasticity.

% $$
% \boldsymbol{c} \equiv\left\langle\overrightarrow{\boldsymbol{Q}} \overrightarrow{\boldsymbol{Q}}^{\mathrm{T}}\right\rangle=\int \overrightarrow{\boldsymbol{Q}} \overrightarrow{\boldsymbol{Q}}^{\mathrm{T}} \psi(\overrightarrow{\boldsymbol{Q}}, t) d \overrightarrow{\boldsymbol{Q}},
% $$

%\section{B. Numerical method and code implementation}
The fluid governing equations (2.3)-(2.5) are solved using the fractional step method on a staggered grid, in a domain of length $L_{d} = 2\pi$ which is discretised into a uniform Eulerian grid with $500^{3}$ cells, which ensures that the ratio between the Kolmogorov dissipative length scale and \textcolor{black}{the grid spacing is $\eta_{0} / \Delta x$ $\approx 0.5$. The adequacy of the grid resolution has also been tested by comparing with results of a $1024^{3}$ cell grid in the single phase flow.} The number of lagrangian points $nl$ on the fiber of length $c$ are decided such that the spatial resolution $\Delta s=c /\left(nl-1\right)$ is approximately equal to the Eulerian grid spacing $\Delta x$. The in-house developed solver Fujin (https://groups.oist.jp/cffu/code), extensively validated in a variety of problems including fibers dispersed in turbulent flows \citep{olivieri2022fully}, has been used. The solver is parallelized using MPI and the 2decomp library for domain decomposition. Second-order central finite-differencing is used for spatial discretisation and the Adam Bashforth scheme for temporal discretisation. \textcolor{black}{Table~\ref{tab:kd} summarises the values of all the parameters used in the study.} %which is based on the (second-order) central finite-difference method for the spatial discretization and the (third-order) Runge-Kutta scheme for the temporal discretization.
\textcolor{black}{We first run the single phase configuration without fibers till obtaining a statistically steady state, which was then used to run a non-Newtonian simulation without fibers. Finally, the obtained flow-fields were used as initial conditions to run the fiber-fluid cases, where the fibers are released randomly in the domain.}

The mutual interactions between the solid and fluid phases are enforced via singular force distributions acting on the fiber and flow, implemented in the setting of an immersed boundary (IB) method \citep{huang2007simulation,rosti2018flexible}. The material points of the immersed fiber are forced to move with the fluid velocities at those points through a no-slip condition $\boldsymbol{\dot{X}} = \mathbf{u}(\mathbf{X}, t)$, where $\mathbf{X} = \mathbf{X}(s, t)$ is the position of a fiber material point and $\mathbf{u}=\mathbf{u}(\mathbf{x}, t)$ is the fluid velocity field. The fluid velocity at the position of the fiber Lagrangian point, $\mathbf{U}=\mathbf{u}(\mathbf{X}(s, t), t)$, is obtained by interpolating the fluid velocity at the Eulerian nodes around the Lagrangian point as:
\begin{equation}    
\mathbf{U}(\mathbf{X}(s, t), t)=\int \mathbf{u}(\mathbf{x}, t) \delta(\mathbf{x}-\mathbf{X}(s, t)) \mathrm{d}^{3} \mathbf{x},
\end{equation}
where the $\delta$ is the Dirac delta function. The interpolated velocity $\mathbf{U}$ is used to compute the fluid-structure forcing term, represented as
\begin{equation}
\mathbf{F}(s, t)=\Upsilon(\mathbf{U}-\dot{\mathbf{X}}),
\end{equation}
where $\Upsilon$ is a large negative constant \citep{goldstein1993modeling,huang2007simulation} \textcolor{black}{with value $-10$}. Finally, the forcing on the fluid from the fiber is calculated as
\begin{equation}
\mathbf{f}_{\mathrm{fib}}(\mathbf{x}, t)=\frac{1}{\rho} \int_{s} \mathbf{F}(s, t) \delta(\mathbf{x}-\mathbf{X}(s, t)) \mathrm{d} s .
\end{equation}

\textcolor{black}{In order to examine if there is a spatial inhomogenity or clustering due to the fibers, the probability density functions of the distance between fibers was computed at a low $De$ and from an instantaneous snapshot. It was seen that the fibers had a distance among each other of approximately $10$ fiber lengths. %were on an average distant several at a large distance between each other compared to their initial length.
%$\approx 0.5$ times distant in comparison to the maximum distance possible in a cubic domain of length $2 \pi$ and also at a large distance between each other compared to their initial length.
This implies that, the fibers are distributed in a homogeneous manner in the domain owing to their low volume fraction, and the chances of spatial clustering are rare and, hence those effects are not being considered in further analysis here.}

% \begin{table}
%   \begin{center}
% \def~{\hphantom{0}}
% \Rotatebox{0}{
%   \begin{tabular}{lccc}
%       $Parameter$  &  Value \\..& ..\\[3pt]
%        $L_{d}$ & $2 \pi$  \\
%        $Re_{\lambda}$& 310 \\
%        $U_{rms}$&6.6 \\
%        $L$&$\approx$ 2.5 \\
%        $\eta_{0}$& 0.005 \\
%         $A=B=C$&5\\
%        $\gamma$ & $[10^{-8},10^{-2}]$ \\
%        $c$ & 0.3 \\
%       $N$&14\\
%       $d$&$ \approx 3 \Delta x$\\
%       $\rho_{l}$& $10^{-3},1$\\
%       $\beta$&0.9\\
%       $\Lambda$&0.3/3, 0.3, 0.3*9\\
%       $nl$&25\\
%       $\Upsilon$&-10\\
      
%   \end{tabular}
%   }
%   \caption{Values of the parameters used in the study.}
%   \label{tab:kd}
%   \end{center}
% \end{table}
\begin{table}
  \begin{center}
  \begin{tabular}{|c|c|c|c|c|c|c|c|c|c|c|c|c|c|c|}
       $L_{d}$ & $Re_{\lambda_{0}}$ & $U_{rms_{0}}$ & $\epsilon_{0}$ &$\eta_{0}$ &$L_{0}$  &$\rho$ &$\nu$& $\beta$ &$\Lambda$ &$N$ &$c$&$d$ &$\gamma$ & $\rho_{l}$\\
      \hline
       $2 \pi$   & 310 & 6.6 & 62&0.0068&2.5& 1& 0.005 & 0.9 & $0.1, 0.3, 2.7$ & 14& 0.3&$\approx 3 \Delta x$&$[10^{-8},10^{-1}]$ & $10^{-3},1$\\
              %  &  &  & & & & &  & & & & 0.3, 2.7 & & &&$10^{-2}]$ &1 & 
       % 0.55  & 1.39128 & ~~1.391 & 1.39131  & 0.1\\
       % 0.7   & 1.32281 & ~10.322 & 1.32288  & 0.1\\
       % 0.913 & 1.34479 & 100.351 & 1.35185  & 0.1\\
  \end{tabular}
  \caption{\textcolor{black}{Values of parameters used in the study:  the domain length $L_d$, the Taylor Scale Reynolds number $Re_{\lambda_{0}}$, the rms velocity $U_{rms_{0}}$, dissipation rate $\epsilon_{0}$, Kolmogorov length scale $\eta_{0}$ and the integral length scale $L_{0}$ of single phase flow, density $\rho$ and kinematic viscosity $\nu$ of the fluid, the solvent to total viscosity ratio $\beta$, the polymer relaxation time $\Lambda$, the number of fibers $N$, the fiber length $c$ and diameter $d$, their bending stiffness $\gamma$, linear density difference $\rho_{l}$. The range of $\gamma$ is spanned in logarithmically equispaced steps.}}
  \label{tab:kd}
  \end{center}
\end{table}

\section{Results}\label{sec:Figures_Tables}

\subsection{Flow dynamics}
%\subsubsection{Energy spectra}
We start the analysis by plotting the turbulent \textcolor{black}{ kinetic energy $E$ (normalized with $\epsilon_{0}^{2/3} \eta_{0} ^{5/3}$, where $\epsilon_{0}$ is the turbulent dissipation rate of the single phase flow), as a function of the wavenumber $k/k_{\eta_{0}}$ ($k_{\eta_{0}}$ being the wavenumber corresponding to the Kolmogorov length scale defined as $2 \pi/\eta_{0}$)} at three different Deborah numbers, $De$ $\approx$ 0.3, 1, 7, to identify if viscoelasticity has played a role in bringing deviations to the classical Kolmogorov spectrum where $E(k)\propto k^{p}$ with $p=-5/3$. It is seen that at the lowest $De$ ($\approx$ 0.3), the spectrum follows the Kolmogorov scaling, as the non-Newtonian effects are weak. As $De$ increases to 1, the scaling behavior is modified: an increasing amount of energy is transferred by the elasticity of the polymers, which alters the spectra to achieve a different scaling coefficient $p= -2.3$. Such a scaling behavior has been already elucidated in non-Newtonian flows without dispersed particles and has been addressed as an `elastic scaling' in experimental and numerical studies \citep {zhang2021experimental,rosti2023large}. The range of $k$ over which the scaling is valid is called the `elastic' range and this case turns out to be a clear case of the interaction of elastic and inertial turbulence. This modified scaling doesn't persist with a further increase in the Deborah number, and the flow recovers the classic Kolmogorov scaling, as shown for $De \approx 7$. Thus, the spectral character of the fluid shows a non-monotonic trend in the presence of polymers, confirming what has been identified in the literature earlier \citep{rosti2023large}. \textcolor{black}{A possible reason for this behaviour is as follows \citep{singh2023bridging}: The polymeric term $\frac{(1-\beta)\eta_{0}}{\lambda}\nabla \cdot \textbf{C}$ , is close to 0 at low $De$, as the polymers stretch minimal and the K41 scaling persists. At $De \approx 1$, the time scales of polymer and flow become comparable, additionally $\nabla \cdot \textbf{C}$ also becomes large as the polymers stretch more, hence a multi-scaling spectrum occurs. At higher Deborah numbers, the polymers, with very large relaxation time ($\Lambda \rightarrow \infty$) cannot follow the carrier flow, or are rather decoupled from the flow, resulting in a return of the Newtonian scaling. In this study, this behaviour is also present since the flow is dilute, and fiber-induced modulation is negligible.} In other words, fibers can be expected to be mere carriers of the information pertaining to the flow, and can possibly be reflective of the polymeric \textcolor{black}{influence} in the flow, if any. The inset of the figure shows $Re_{\lambda}$ (\textcolor{black}{based on the fluid dissipation rate and rms velocity in the polymeric flow)} for each of the three Deborah numbers, and clearly is seen to increase with respect to the single-phase flow (\textcolor{black}{shown with dotted lines}), as the fluid dissipation rate drops due to the presence of polymers. 
\begin{figure}
\centering
\includegraphics[width=1\textwidth,,trim={2cm 0cm 4cm 1cm},clip]{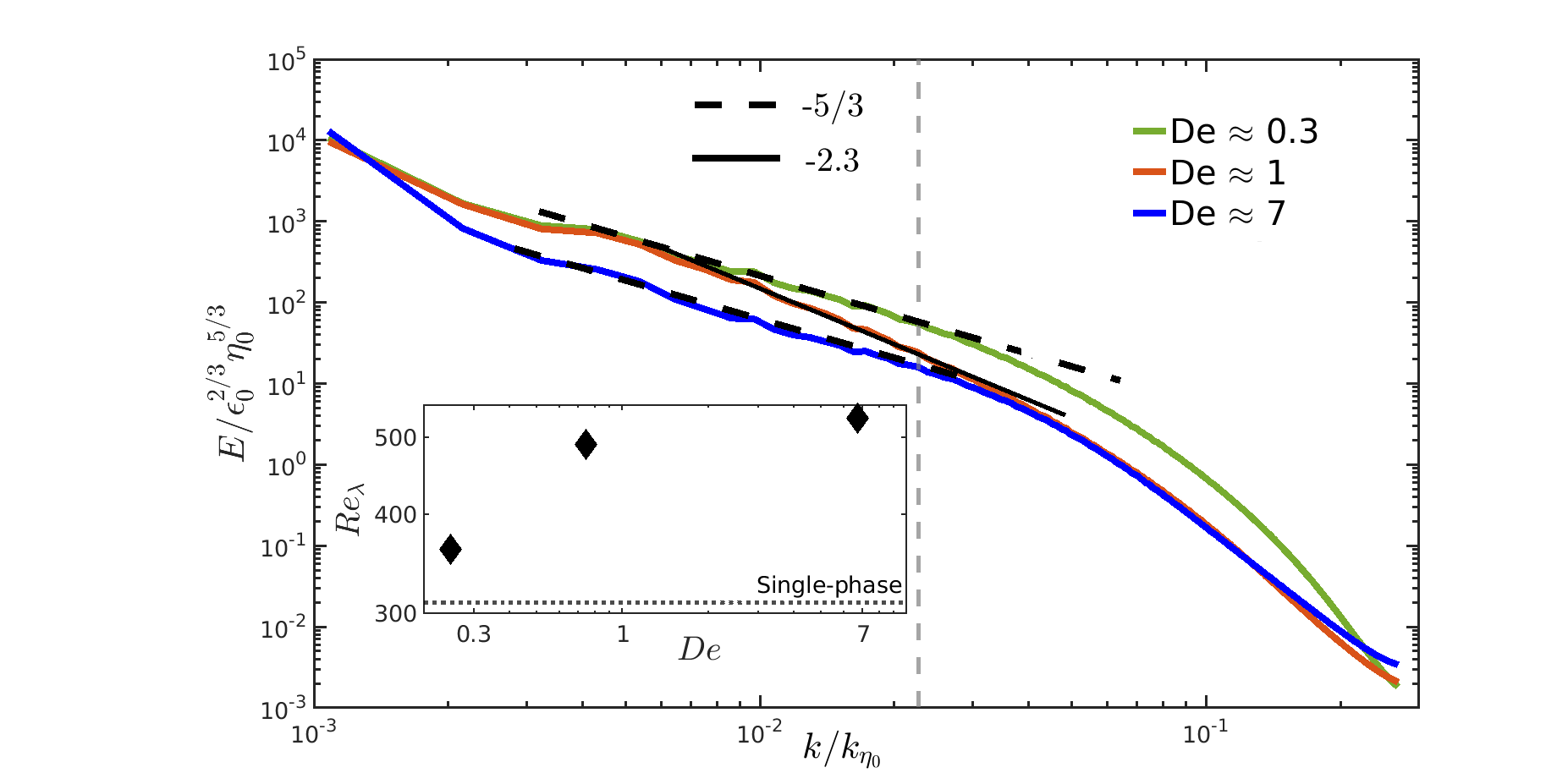}% Here is how to import EPS art
\label{fig:Espec}
\caption{Energy spectra of the turbulent polymeric flows at different Deborah numbers $De$. The vertical dashed line represents the wavenumber corresponding to the fiber length. The inset shows the resulting Taylor Reynolds number $Re_{\lambda}$ for each of the cases.}
\end{figure}

\begin{figure}
%\centering
\begin{subfigure}[b]{0.5\textwidth}
\centering
\includegraphics[width=1\textwidth,trim={1cm 0cm 4cm 0cm},clip]{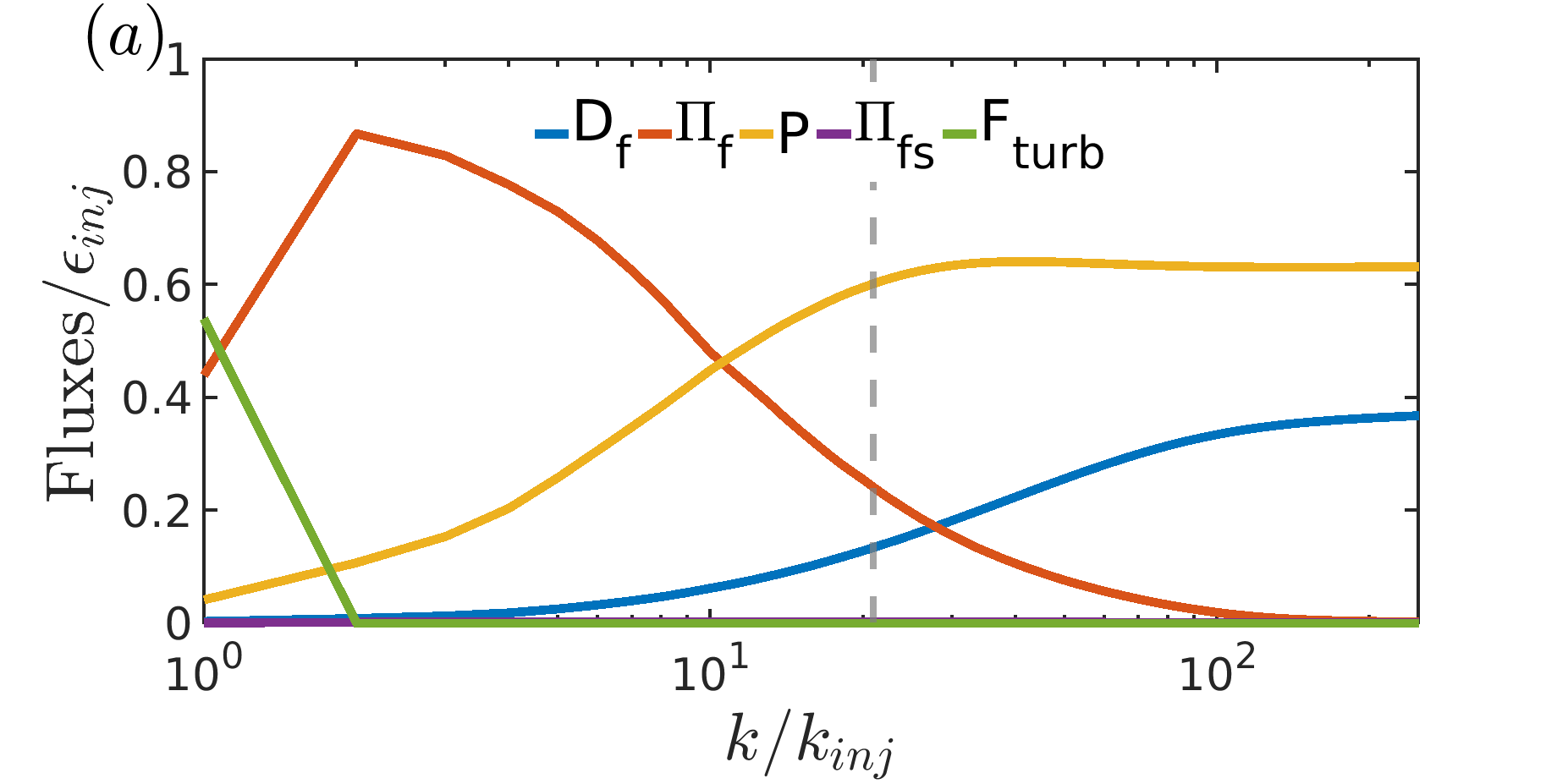}% Here is how to import EPS art
%\caption{}
\end{subfigure}
\hfill
\begin{subfigure}[b]{0.5\textwidth}
\centering
\includegraphics[width=1\textwidth,trim={1cm 0cm 4cm 0cm},clip]{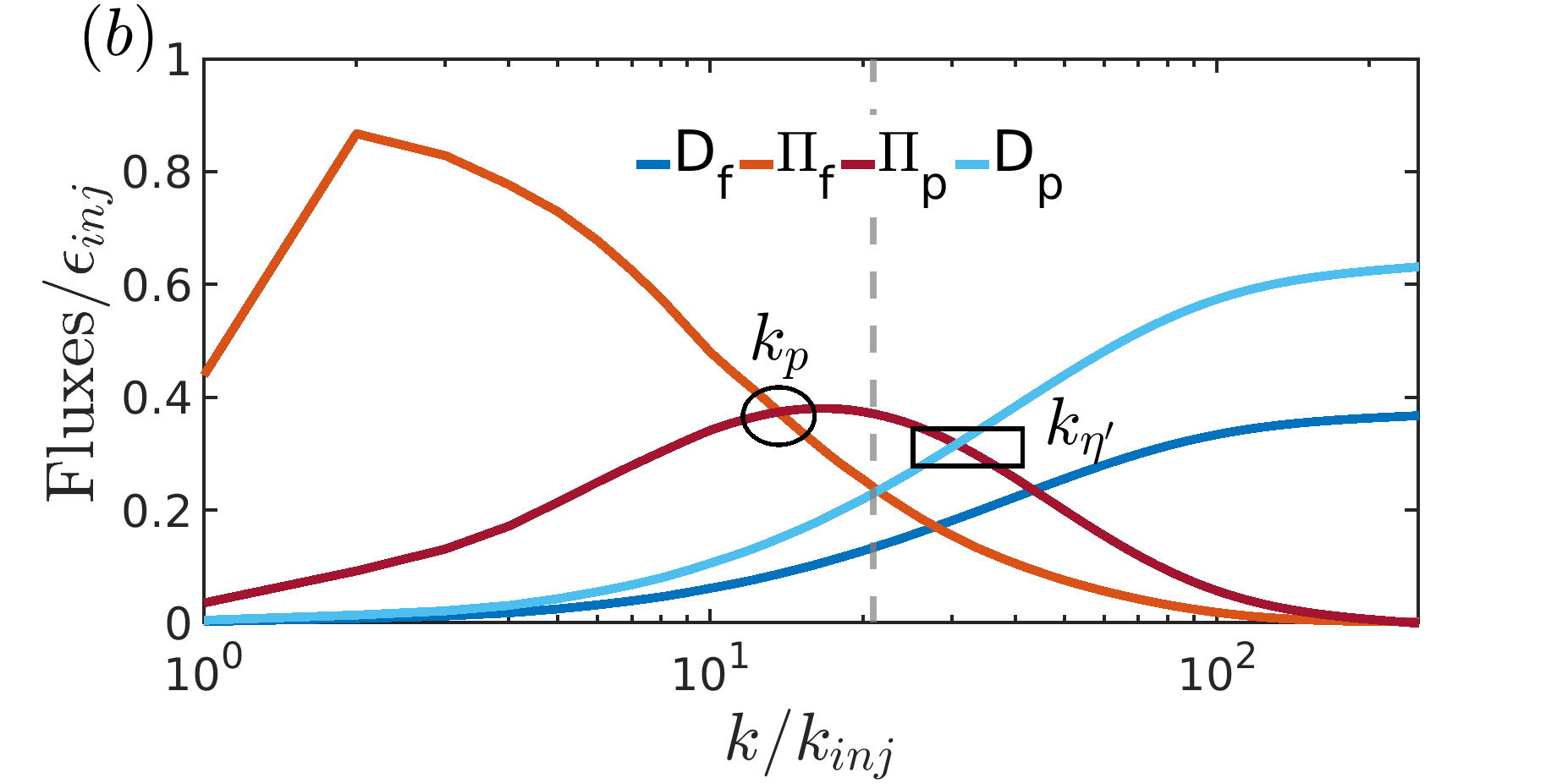}% Here is how to import EPS art
%\caption{}
\end{subfigure}
\caption{(a)  Flux contributions from Eq \ref{eqfluc} (normalized with the turbulent energy dissipation rate) plotted with respect to the wave number $k$, at $De \approx 1$. (b) The variation of the nonlinear energy flux $\Pi_{f}$, fluid dissipation $D_{f}$, polymer flux $\Pi_{p}$ and the polymeric dissipation $D_{p}$. The vertical dashed lines represent the wavenumber corresponding to the fiber length. }
\label{fluxab}
\end{figure}

Following this, we perform a scale-by-scale budget analysis at the intermediate Deborah number $De \approx 1$ to obtain the flux of kinetic energy
%\citep{rosti2023large,olivieri2022fully}:
\begin{equation}
    \epsilon_{inj}=\epsilon_{f}+\epsilon_{p}=\Pi_{f}(k)+D_{f}(k)+P(k)+\Pi_{fs}(k)+F_{turb}(k).
    \label{eqfluc}
\end{equation}
Here, $\epsilon_{inj}$ is the total injected dissipation rate, $\epsilon_f$, and $\epsilon_p$ are the corresponding components from the fluid and the polymer; $\Pi_{f}$, $D_{f}$, $P$, $\Pi_{fs}$, and $F_{turb}$ are the flux contributions from the nonlinear term, viscous term, polymeric stresses, the fluid-structure coupling, and the external forcing, respectively. The variation of each of the above flux components (normalized with $\epsilon_{inj}$),  as a function of wave number \textcolor{black}{(normalised with $k_{inj}$, the forcing wavenumber)} is shown in Fig \ref{fluxab}a. An overall observation shows that the contribution from the forcing $F_{turb}$ prevails only at the lowest wave number and that the fluid-structure coupling $\Pi_{fs}$ is negligible as the volume fraction of the fibers is very low. The nonlinear term $\Pi$ dominates at low wave numbers, taken over by the fluid dissipation $D_{f}$ at higher $k$s, which eventually saturates to $\epsilon_{f}$. However, at this Deborah number, the effects from the polymer stresses $P$ dominate over $D_{f}$, as one of the effects of polymers is to increase the effective dissipation \citep{hinch1977mechanical,lumley1973drag,bird1987dynamics}. However, $P$ is not a purely dissipative term, evident through its non-monotonic behavior with $k$. \cite{rosti2023large} proposed to break down $P$ as: 
 \begin{equation}
P(k)=\Pi_{p}(k)+D_{p}(k),
\end{equation}
where
\begin{equation}
D_{p}(k)=\frac{\epsilon_{p}}{\epsilon_{f}}  D_{f}(k).
 \end{equation}   
Here, $D_{p}$ is a purely dissipative component that saturates to the polymeric dissipation rate $\epsilon_{p}$, and $\Pi_{p}$ a nonlinear component from the polymer contribution alone. Further, they evaluated the range over which the elastic scaling is valid by simultaneously plotting $D_{f}$, $\Pi_{f}$ and $\Pi_{p}$, which for the present case is shown in Fig \ref{fluxab}b. At low $k$, the flux is carried predominantly by the solvent $\Pi_{f}$ which diminishes as $k$ increases, while the polymeric flux $\Pi_{p}$ increases. The crossover $k$ between these two fluxes is identified as $k_{p}$ and the crossover between $\Pi_{p}$ and $D_{p}$ (or $D_{f}$) is defined as $k_{\eta'}$, which is the wave number when any of the dissipation dominates \citep{rosti2023large}. The intermediate range $k_{p} < k < k_{\eta'}$ is the elastic range. This exercise has been done in the current context to show that the wave number corresponding to the fiber length scale $c$, equal to $k_{c}/k_{inj} \approx 21 $ \textcolor{black}{(corresponding to the dashed vertical line in Figs 2, 3 at $k/k_{\eta_{0}}$ $\approx 0.02)$} falls within this elastic regime. In other words, the fiber is in a range of length scales that is dominated by the fluid elasticity. %The fluxes clearly illustrate and substantiate what we already observed in the energy spectra: For k < kp, the turbulence is Kolmogorov-like, whereas for kp < k < kη, the polymeric flux Πp dominates and is approximately a constant. We define the range of Fourier modes, kp < k < kη as the elastic range with kp precisely defined as Πf(kp) = Πp(kp).

\subsection{Fiber dynamics}
% We analyse the flapping dynamics separately for the neutrally-bouyant as well as the denser-than-fluid fibres separately in this section. Figure \ref{}shows the power of the fibre frequencies plotted at different the three different Deborahs. Primarily, the fibre response is charactetrised by three different frequency components (i) from the forcing to the flow, represented by the orange region (ii) a frequency close to 1, which is also approcimately the integral frequency of the systm, and (iii) higher frequencies in (2.5,8-10). However, the way in which these frequencies are trigerred is very different based on Deborahs and based on the linear density differences. These are discussed as follows.
\subsubsection{Flapping frequency}

We now turn our focus into identifying the dominant flapping dynamics of the fibers across different Deborah numbers, by primarily looking at their frequency of flapping. The flapping frequency of fibers interacting with fluids is indeed a well-probed quantity in previous works, due to the interesting transitions it shows with respect to various parametric variations. Notable mentions in this context are the works by \cite{rosti2020flowing} and \cite{olivieri2022fully}, wherein the potential of finite-size flexible fibers to measure relevant two-point statistics of turbulence was highlighted. Mainly, two qualitatively different dynamical regimes were identified: \textit{(i)} one controlled predominantly by the flow time scales, with the fibers acting as a representative of the turbulent flow; \textit{(ii)} another controlled by the fiber's natural frequency, in which the effects coming from the carrier flow are negligible. Extending this analysis in the context of a viscoelastic flow scenario for a dilute configuration of dispersed particles is one of the main foci of this study. We perform a detailed analysis to evaluate the frequency content of the fiber response using a continuous time wavelet transform and Fast Fourier Transform (FFT) on the end-to-end displacement ($y$) of fibers with different bending rigidities $\gamma$, at various Deborah numbers. %The objectives are two-fold: \textit{(i)} to identify how the fiber flapping frequencies evolve with respect to $\gamma$, at different $\rho_{l}$ and $De$; \textit{(ii)} to identify the effect of $D$ itself. 
A goal of this work is to recognize if the fiber at some/all parametric combinations is reflective of changes in the fluid due to the presence of polymers. In this context, it is worth mentioning certain important frequency values associated with the system: \textit{(i)} the large eddy-turnover frequency $f_{l}$ = $U_{rms_{0}}$/${L_{0}}$, where $U_{rms_{0}}$ is the root mean square velocity associated with the single phase flow and $L$ is the integral length scale; \textit{(ii)} the eddy-frequency at the fiber's length scale $f_{c} \approx c^{-2/3} \epsilon^{1/3}$, where $\epsilon$ is the turbulent dissipation rate of kinetic energy and the formula holds for Newtonian fluids; \textit{(iii)} the frequency associated with the polymer stretching, $f_\Lambda$ = $1/\Lambda$%= 10, 3, 0.3 at $De$ $\approx$ 0.3, 1, 7, respectively
; and finally, \textit{(iv)} the fiber natural frequency (from a normal mode analysis of Eq 2.1) given by $f_{\text {nat }}=\alpha \sqrt{\gamma /\left({\rho_{l}} c^{4}\right)}$ (where $\left.\alpha \approx 22.4 / \pi\right)$.

\begin{figure}
    \centering
    \includegraphics[width=1.05\textwidth,trim={2cm 0 2cm 0cm},clip]{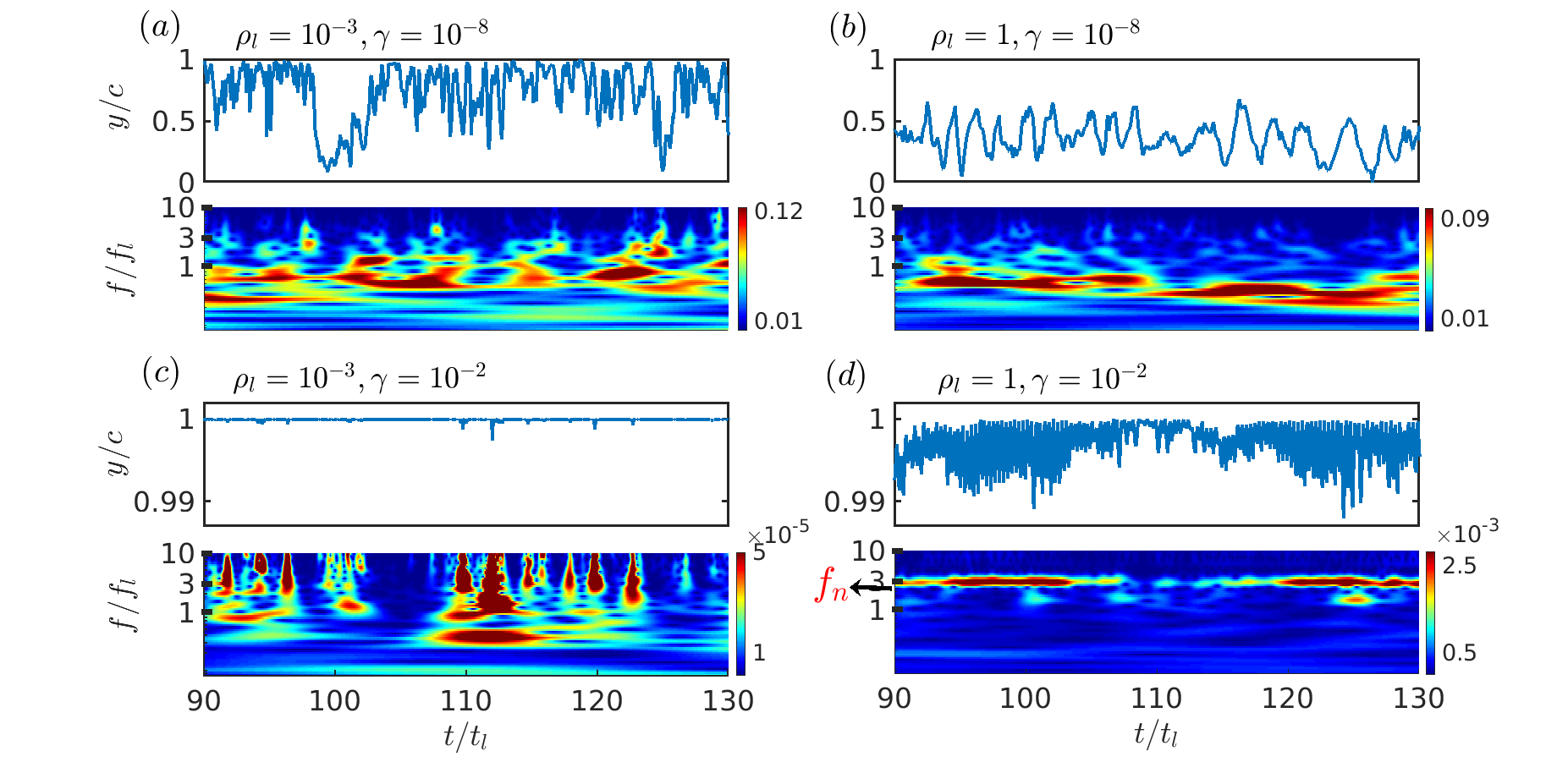}
    \caption{Time histories and Wavelet transform of the fiber end-end displacement at the smallest Deborah number $De \approx 0.3$ for (a) neutrally bouyant, flexible ($ \rho_{l} = 1 0^{-3},\gamma = 10^{-8}$), (b) denser-than-fluid, flexible ($\rho_{l} =1,\gamma = 10^{-8}$), (c) neutrally bouyant, rigid ($\rho_{l}=10^{-3},\gamma=10^{-2}$), (d) denser-than-fluid, rigid ($\rho_{l} = 1,\gamma = 10^{-2}$) fibers.}
    \label{fig:thwv}
\end{figure}
The time histories of the end-to-end displacement $y$ normalized with the fiber unbent length $c$ are used to explore its frequency behaviors. Note that FFTs are useful in capturing the global frequency features of the responses in contrast to wavelets which are time-frequency transforms that help analyze the local characteristics in time. Hence, a complimentary analysis based on both is useful in the analysis of such highly non-stationary signals and thus such an approach has been used in this work to analyze the flapping frequencies. Fig \ref{fig:thwv} discusses the time histories and their wavelets for two extreme values of rigidities, at $\gamma$ = $10^{-8}$ (flexible) and $\gamma$ = $10^{-2}$ (rigid) (top and bottom). The left and right panels show the cases at two different values of linear densities, $\rho_{l}$ = $10^{-3}$ and ${\rho_{l}}$ = $1$. The former represents an approximately `neutrally-bouyant' scenario and the latter corresponds to `denser than the fluid' fibers and shall be addressed so hereon in the manuscript. As a starting point, we discuss the flexible fiber: left-top case and right-top case (Fig \ref{fig:thwv} a,b). Clearly, the two cases show definite differences: the time histories of neutrally-bouyant fibers show rapid variations, intermittently fluctuating between amplitudes as low as $0.1$ to amplitudes as high as the fiber's initial length itself ($y/c = 1$). On the contrary, the fluctuations are more bounded around similar amplitudes for the denser fiber, with a mean value around $0.4$. With respect to the frequency content of the signal, at low $\gamma$, it can be seen that the neutrally-bouyant responses are dominantly characterized by a broad spectrum of frequencies approximately around $f/f_{l} \approx 0.05-3$ (Fig \ref{fig:thwv} left-top), whereas the denser fiber shows a relatively narrower spread of frequency values around or below $f/f_{l}$  $\approx $ 1 (Fig \ref{fig:thwv} right-top). Note that, frequencies are normalized with the large eddy turnover frequency $f_l$.

For the rigid fiber, (Fig \ref{fig:thwv} c, d ) corresponding to neutrally-bouyant (left-bottom) and denser fibers (right-bottom) respectively, the time histories convey that fiber tends to remain in a more unbent configuration, and the frequency behaviors vary drastically across the two linear densities. As the fiber gets more rigid, it starts flapping with higher frequencies, see e.g. Fig \ref{fig:thwv}c where peaks appear up to $f/f_{l}$ $\approx$ 10. \textcolor{black}{This may be caused by the different shape of the fiber itself. Rigid fibers deform less, or are closer to the initial unbent configuration, and thus can react to the small-scale as well as large scale flow structures around them, resulting in small as well as large time scales in the fiber spectrum.} The rigid, denser one (Fig \ref{fig:thwv}d) shows a band of frequency at around $f/f_{l} \approx 3$, which corresponds to the natural frequency $f_n/f_{l}$ at this $\gamma$. The four different cases compared in Fig \ref{fig:thwv} suggest that the flapping dynamics of the fiber (temporal and spectral) is strongly influenced by its density and rigidity: the neutrally bouyant fibers flap with a broader spectrum of fluid time scales: with very large time scales when they are flexible and a combination of large and smaller time scales as they become more rigid. The denser fibers, however, are limited in their spectrum: with very large time scales related to the flow when they are flexible or the fiber's natural frequency as they become rigid. \textcolor{black}{The above discussion through wavelets were made to develop an understanding of how the fiber flaps at low Deborah number, close to the classical Newtonian scenario. It is in this picture that we attempt to characterise the effects of polymer elasticity on the flapping dynamics of the fiber in the following part of the manuscript.}
 
% \begin{figure}
% %\centering
% \begin{subfigure}[b]{1\textwidth}
% \centering \includegraphics[width=1\textwidth,trim={1cm 0 96 100},clip]{figures/wvt.eps}% Here is how to import EPS art
% \caption{}
% \end{subfigure}
% \begin{subfigure}[b]{1\textwidth}
% \centering %\includegraphics[width=1\textwidth,trim={1cm 0 96 100},clip]{figures/wvt2.eps}% Here is how to import EPS art
% \caption{}
% \end{subfigure}
% \caption{The time histories and wavelets of the end-to-end distance $X$ of the fiber at the lowest $De$ = 0.1, for (a) $\gamma$= $10^{-8}$ (b) $\gamma$= $10^{-2}$. The left and right panel shows $\Delta \tilde \rho$= $10^{-3}$ and $1$ respectively.}
% \label{thwv}
% \end{figure}

\begin{figure}
%\centering
\begin{subfigure}[b]{0.5\textwidth}
\hspace{-0.5 cm}
\includegraphics[width=1.09\textwidth,trim={2cm 0 6 0},clip]{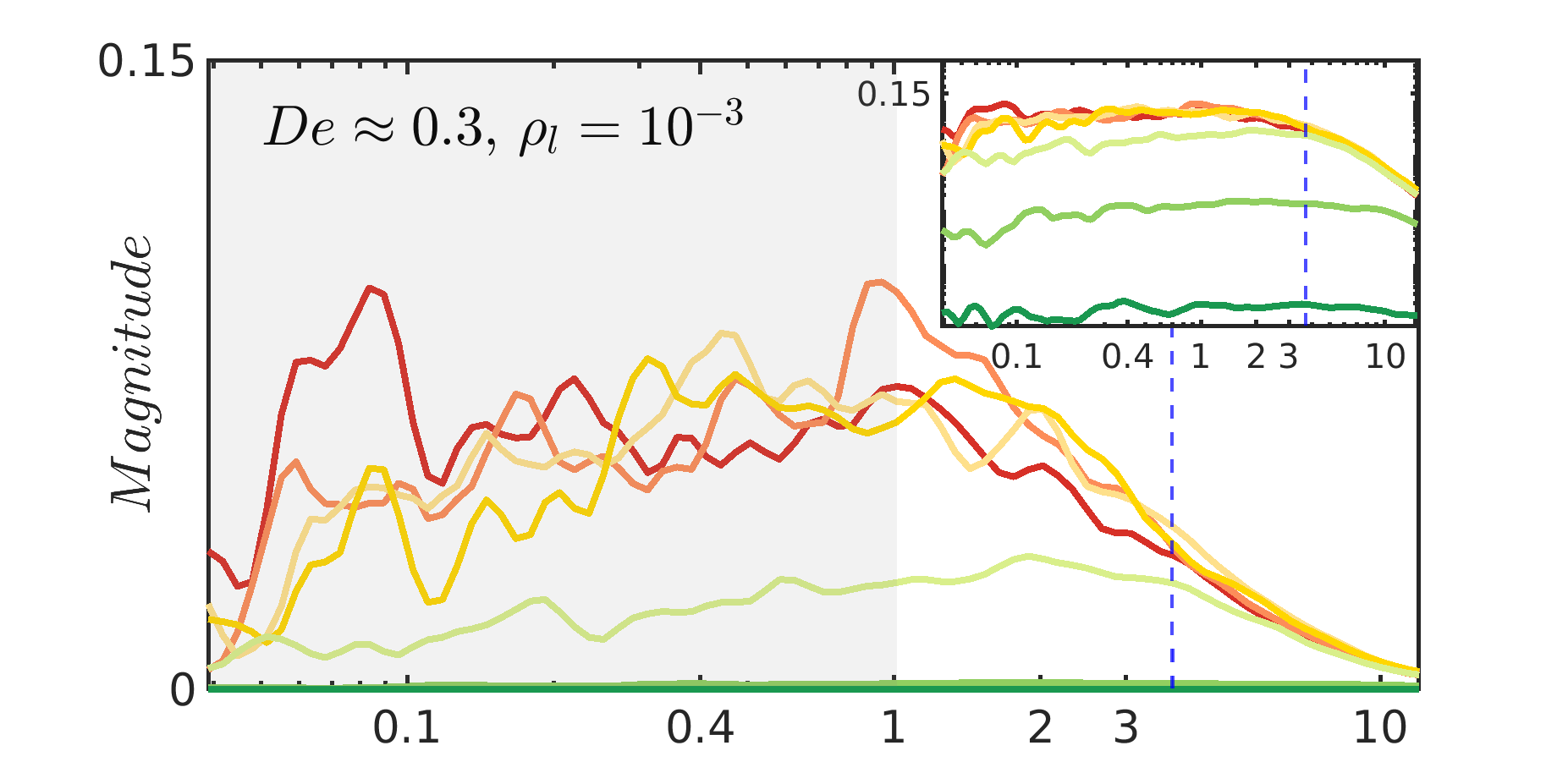}% Here is how to import EPS art
\caption{}
\end{subfigure}
\begin{subfigure}[b]{0.5\textwidth}
\hspace{-0.5cm}
\includegraphics[width=1.09\textwidth,trim={2cm 0 6 0},clip]{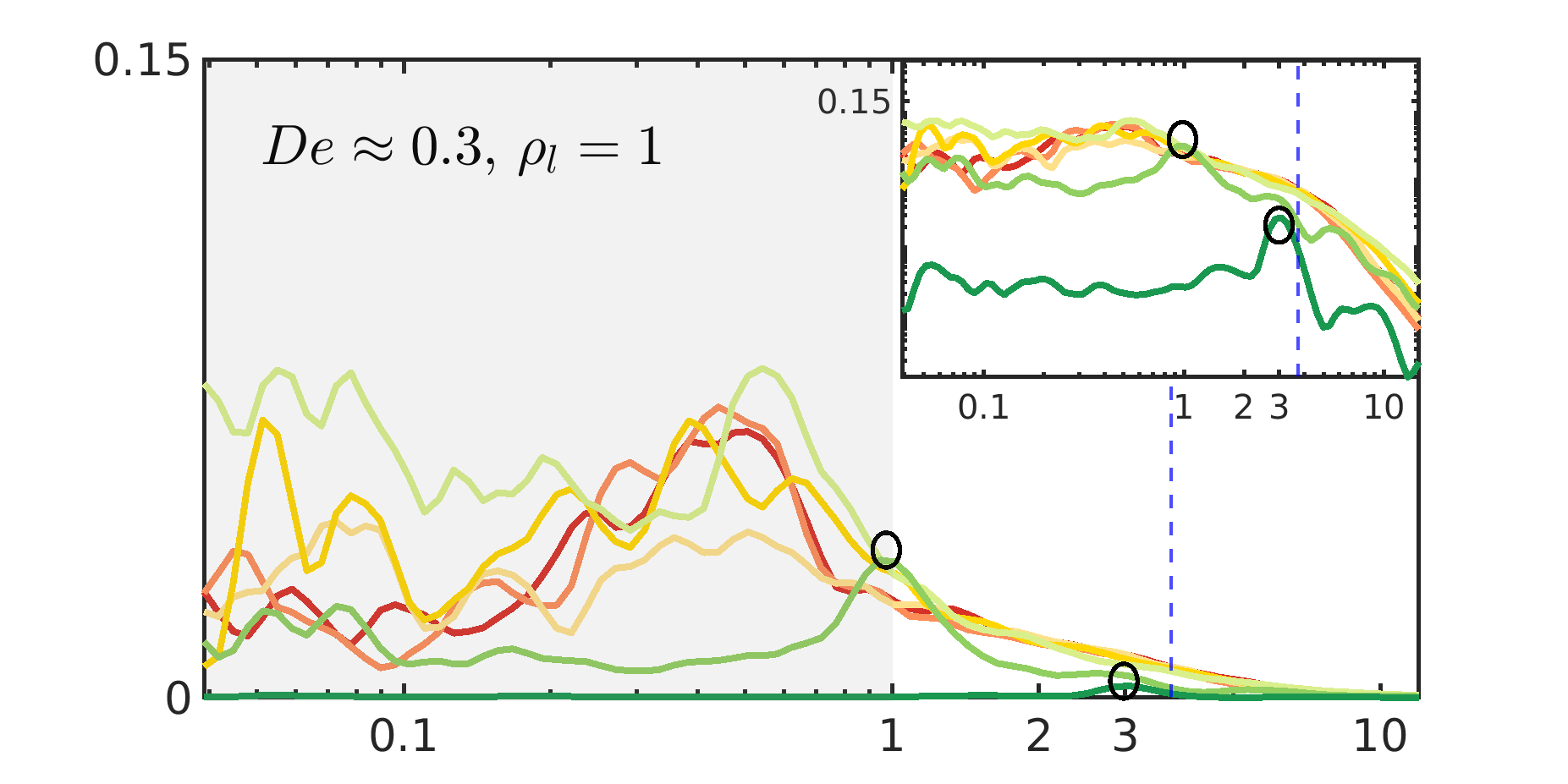}% Here is how to import EPS art
\caption{}
\end{subfigure}
\begin{subfigure}[b]
{0.5\textwidth}
\hspace{-0.5 cm}
\includegraphics[width=1.09\textwidth,trim={2cm 0 6 0},clip]{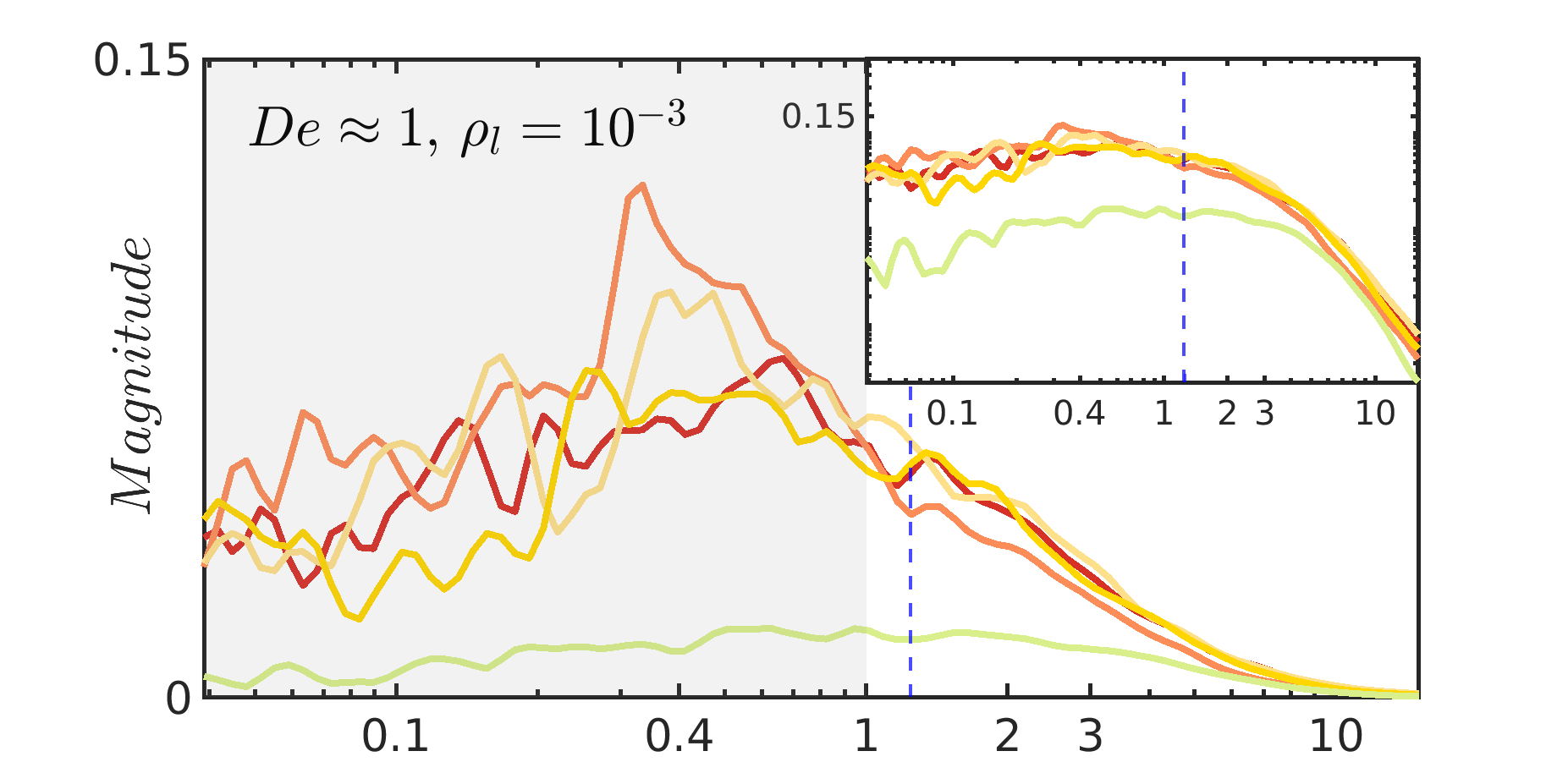}% Here is how to import EPS art
\caption{}
\end{subfigure}
\begin{subfigure}[b]{0.5\textwidth}
\hspace{-0.5 cm}
\includegraphics[width=1.09\textwidth,trim={2cm 0 6 0},clip]{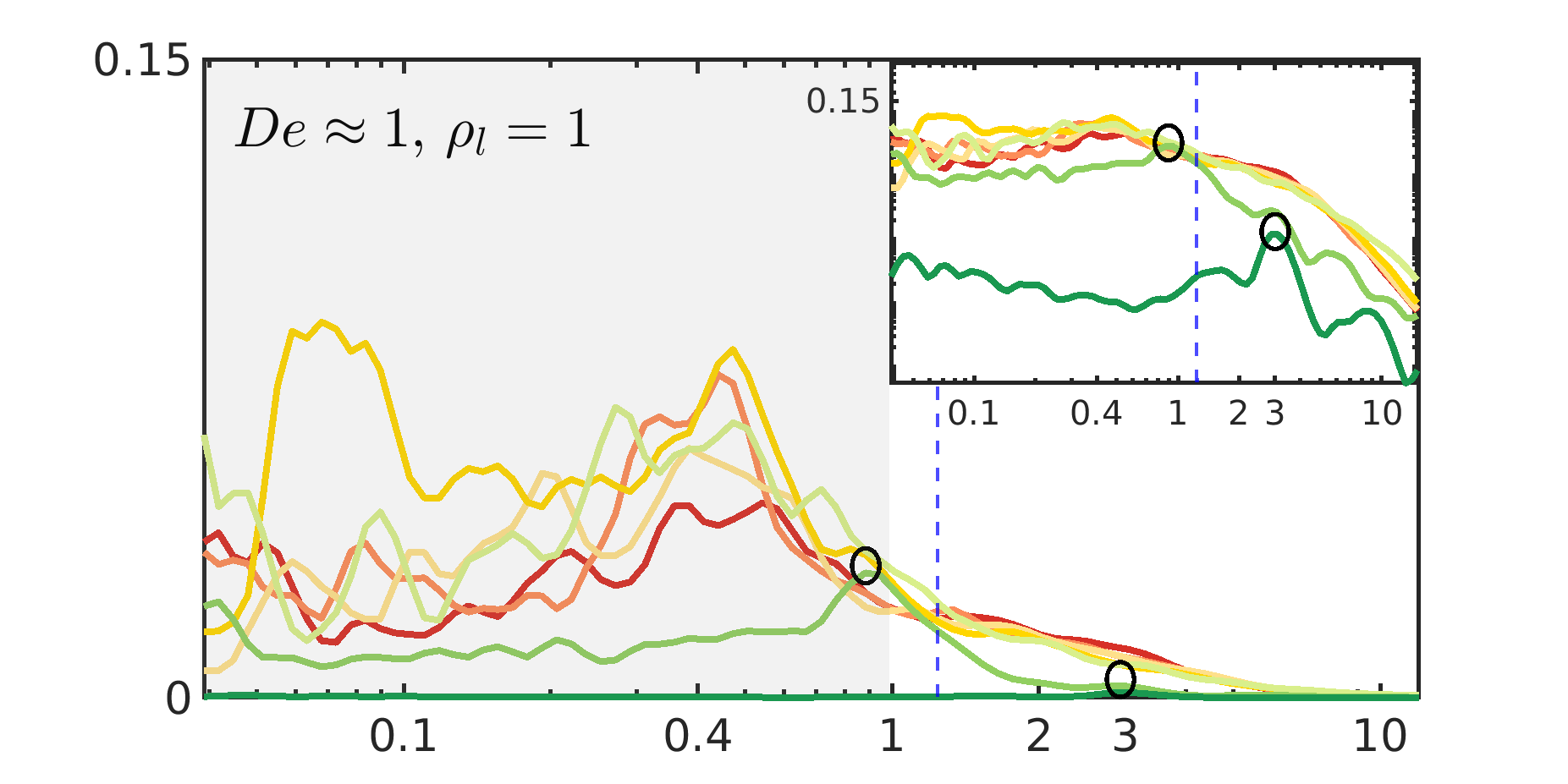}% Here is how to import EPS art
\caption{}
\end{subfigure}
\begin{subfigure}[b]{0.5\textwidth}
\hspace{-0.5 cm}
\includegraphics[width=1.09\textwidth,trim={2cm 0 6 0},clip]{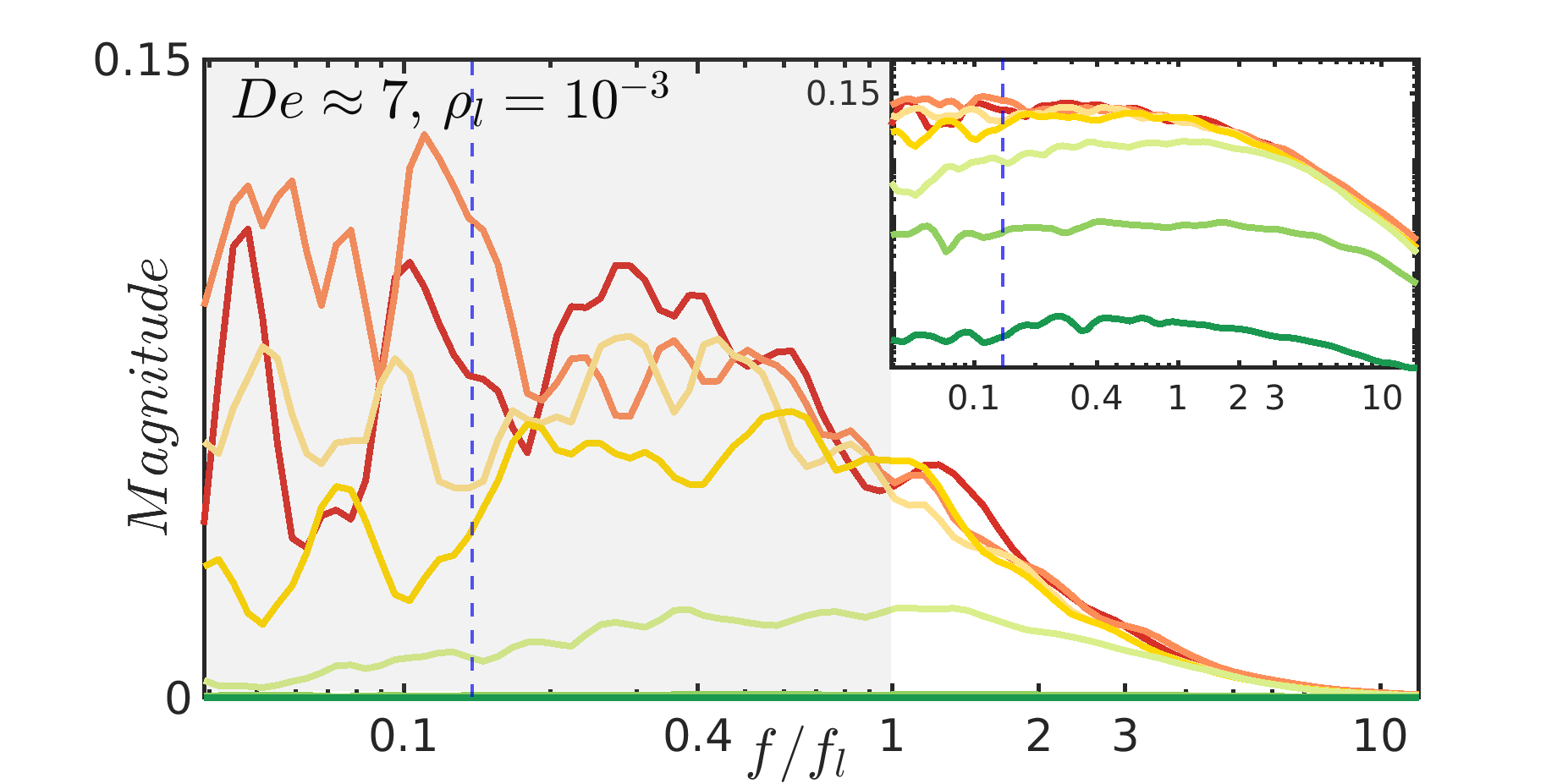}% Here is how to import EPS art
\caption{}
\end{subfigure}
\begin{subfigure}[b]{0.5\textwidth}
\hspace{-0.5 cm}
\includegraphics[width=1.09\textwidth,trim={2cm 0 6 0},clip]{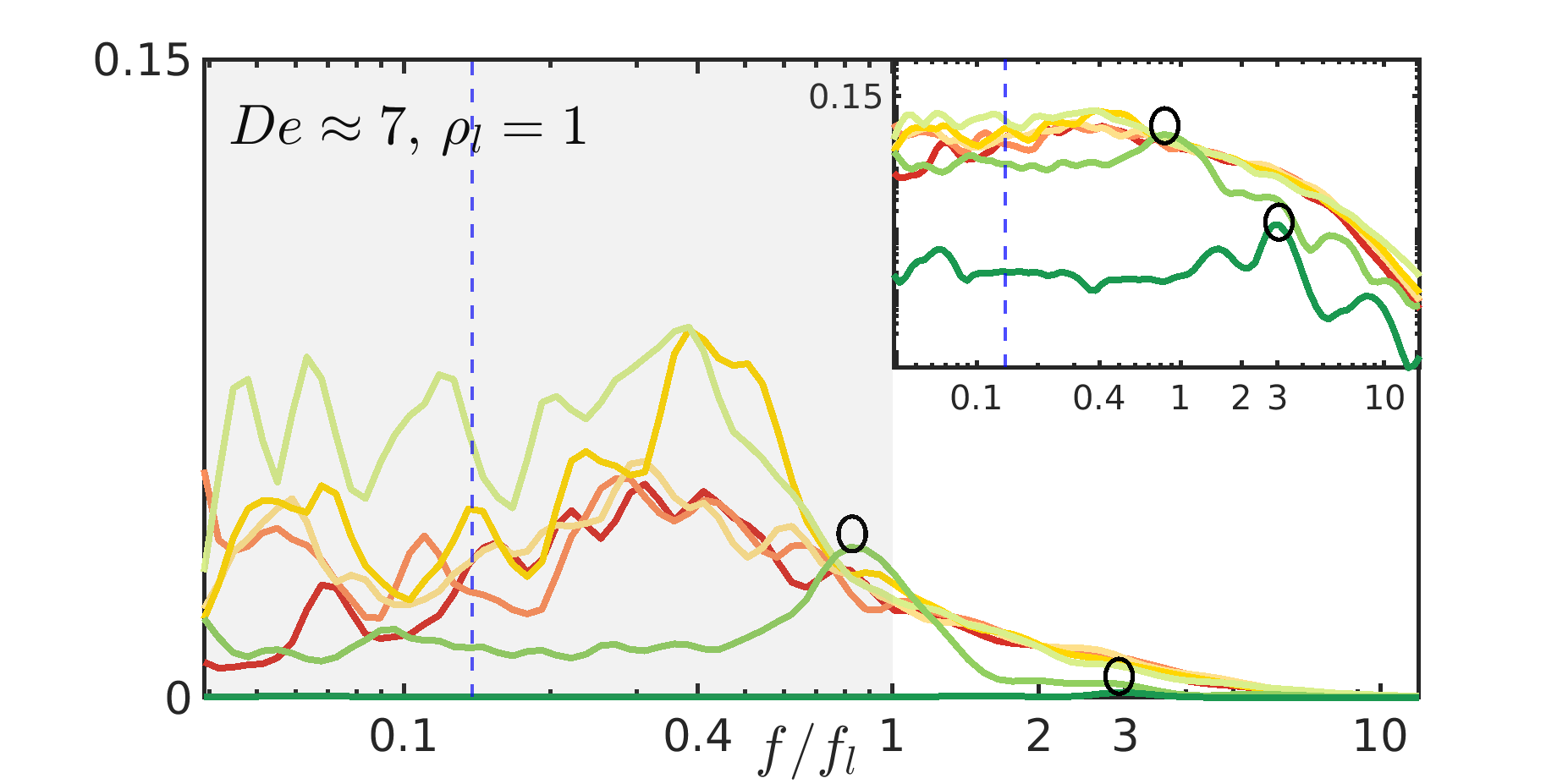}% Here is how to import EPS art
\caption{}
\end{subfigure}
\begin{subfigure}[b]{0.8\textwidth}
\hspace{1 cm}
\includegraphics[width=1\textwidth,trim={0cm 0 0 0},clip]{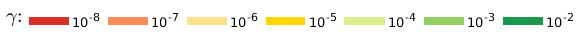}% Here is how to import EPS art
\end{subfigure}
\caption{Frequency distribution of the fibers from the wavelet analysis, for $De \approx 0.3, 1, 7$ (top, middle, bottom rows). The left panels show neutrally-bouyant ($\rho_{l}$ = $10^{-3}$) and the right panel shows denser-than-fluid ($\rho_{l} = 1$) fibers.}
\label{ffmean}
\end{figure}

As a next step, we attempt to confirm the existence of this behavior across different Deborah numbers $De$ in Fig \ref{ffmean}: the abscissa represents the frequencies exhibited by the fiber, and the ordinates show their magnitude. The left and right panels show neutrally-bouyant and denser fibers respectively, and the three rows correspond to Deborah number increasing from top to bottom, with each plot showing the variation in $\gamma$s. To plot each of these figures, a wavelet analysis (as shown earlier) was performed first. Then, the magnitudes of the wavelet transform of each frequency were averaged in time and subsequently plotted as a function of the frequencies in Fig \ref{ffmean}. Such an approach has been used to explain and help visualize the dominant and most relevant frequencies to the system which are otherwise inherently encapsulated by the variety of timescales and noise introduced by the turbulence. Essentially we see the same information as in Fig \ref{fig:thwv}, but with the temporal effects now averaged out.

We classify the frequencies exhibited by the fibers into two different categories, addressed as a low-frequency/large time-scale and high-frequency/small time-scale regime, shaded by gray and white regions in Fig \ref{ffmean}. To visualize specific details of the dynamics, the main plots are represented in a log-linear scale, whereas the insets are shown in a log-log format. The low-frequency regime is due to the large-scale eddies in the flow, while the high-frequency window is from the small-scale eddies or the natural frequency $f_n$ of the fiber itself. We will show now that, the flapping frequency of the fiber is dictated by one of these frequencies in each regime, or a combination of them depending on  $\gamma$, $\rho_{l}$, and $De$.

\begin{figure}
 
\includegraphics[width=1\textwidth,trim={2cm 2.5cm 6 0},clip]{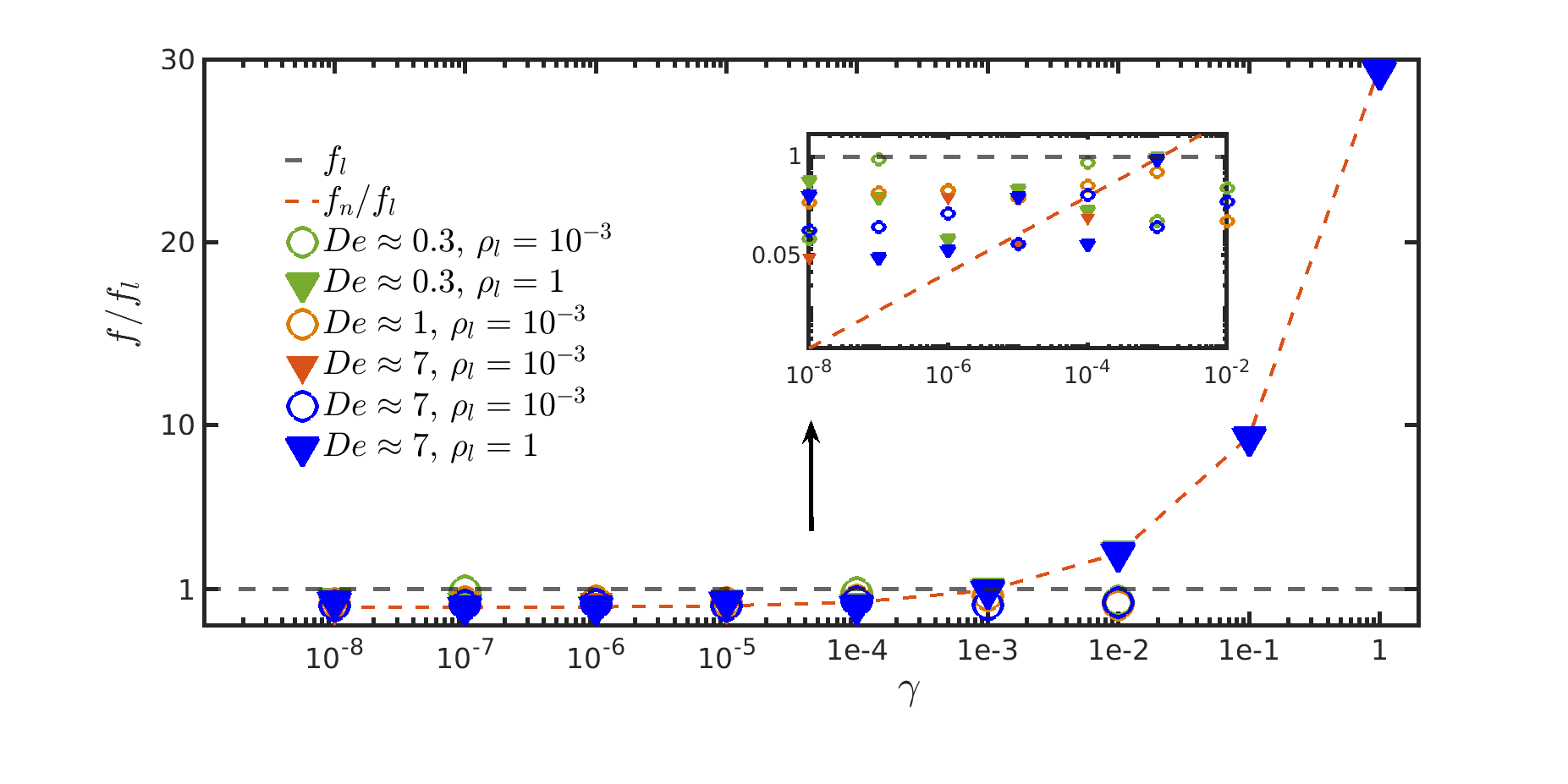}% Here is how to import EPS art
\caption{The flapping frequency of the fiber from a Fourier analysis, as a function of the bending rigidity $\gamma$. The inset reports a zoomed view of the same plot in a log-log scale. \textcolor{black}{For dense fibers, we include results from additional simulation with $\gamma =10^{-1}$ and $1$.}}
\label{fft}
\end{figure}

% \begin{figure}
%   \begin{subfigure}[b]{0.5\textwidth}
% \includegraphics[width=1\textwidth,trim={2cm 0 6 0},clip]{figures/fiso.eps}% Here is how to import EPS art
% \caption{\label{fig:epsart}}
% \end{subfigure}
%  \begin{subfigure}[b]{0.5\textwidth}
% \includegraphics[width=1\textwidth,trim={2cm 0 6 0},clip]{figures/fdenser.eps}% Here is how to import EPS art
% \caption{\label{fig:epsart}}
% \end{subfigure}
% \caption{The flapping frequency of the fiber as a function of the bending rigidity $\gamma$ for (a) $\Delta \rho$=$10^{-3}$ (b)  $\Delta \rho$=$1$}
% \label{fft}
% \end{figure}

First, we examine the dynamics for the neutrally bouyant case (left panel) at all three Deborah numbers. The different colors represent the bending rigidities $\gamma$, with $\gamma $ $\in$ $[10^{-8}, 10^{-2}]$ in steps of $10^{-1}$, roughly increasing from top to bottom. The blue dotted vertical line represents the normalised polymeric frequency $f_{\Lambda}/f_{l}$. Consistent with the previous observations in Fig \ref{fig:thwv}, at the lowest $De$ (Fig \ref{ffmean}a), low $\gamma$ fibers are influenced mainly by the large time scales, with the contribution from the smaller time scales growing as the rigidity increases (see inset). Indeed, at the highest $\gamma$s, high frequency peaks upto $f/f_{l} = 10$ start emerging. Physically this means that the flexible fibers are mainly controlled by the large-scale structures of the flow, and as the fiber becomes more rigid, it starts being affected also by the smaller eddies around it. At the intermediate $De$ $\approx$ 1 (Fig \ref{ffmean}c), the polymeric frequency shifts left as it is an inverse function of $De$. We would like to see if there is a difference in the fiber flapping with this change. The main plot shows that there is not a spread in frequencies like at the lowest $De$: indeed, frequencies are peaking around $f/f_{l}$ $\approx$ 0.4 and the high-frequencies are less evident for the rigid fibers (inset). As $De$ is further increased, and $f_{\Lambda}$ further shifts leftwards (Fig \ref{ffmean}e), a significant reduction in the magnitude of the high frequencies can be seen for both low and high $\gamma$. Overall, there seems to be a subdominant resonating effect between the excited time-scales and the polymeric time-scale, resulting in the dominant fiber flapping peaks shifting to smaller values as the polymeric frequency reduces. In other words, the results suggest that an increase in the polymer relaxation time also suppresses the smaller time scales which are otherwise picked up the fiber. %It is of our interest to assess what can be a possible reason for this subdominant, still non-negligible quantitative shift to happen. 
The effects observed in the Lagrangian spectra of the fiber seem to be correlated to what is already known regarding the role of polymers in influencing the turbulent flow-structures, i.e., that polymers smooth out small scale structures and eddies in the flow, and the same effect is transferred to the dynamics of the fibers. Indeed, as the small-scale structures of the flow are damped with increasing $De$, the fibers flap with larger time-scales. 

Next, we discuss the flapping dynamics of the denser-than-fluid fibers in the right panel (Fig \ref{ffmean}b, d, f). The major differences are: (i) Compared to the neutrally-bouyant case, there is no discernible high-frequency regime at low $\gamma$s, and the fiber primarily flaps with the largest time-scales of the flow, (ii) As $\gamma$ is increased, a singular high-frequency peak is triggered, corresponding to the fiber natural frequency $f_n/f_l$, represented by the black circles in the plots. These features are persistent across all the Deborah numbers, indicating a reduced effect of the polymers on the fiber dynamics as they get more inertial.

We show  a global picture of the frequencies by showing the dominant flapping frequencies obtained from a Fast Fourier Transform of the fluctuation of the end-end displacement, plotted as a function of $\gamma$ in Fig \ref{fft}. \textcolor{black}{Note that, we report in the figure results from additional simulation with $\gamma =10^{-1}$ and $1$ for the denser fibers.} In particular, we plot the average of all the frequencies captured by the Fourier analysis which have at least 90$\%$ prominence as the most dominant peak. The black line in the figure corresponds to the large eddy turnover frequency $f_l$, while the red line to the natural frequency $f_n$, with all the frequencies being normalised with $f_l$. The most important features are that \textit{(i)} there is a flow-driven and a fiber-driven flapping regime, \textit{(ii)} the denser fibers fall into either of these regimes depending on $\gamma$ and show minimum sensitivity with respect to $De$, \textit{(iii)} the neutrally-bouyant fibers always fall in the flow-driven regime, but show more sensitivity to the variation in $De$. Numerical simulations of rigid fibers moving in laminar flows by \cite{cavaiola2020assembly} and in turbulent flows by \cite{brizzolara2021fiber} revealed that, as the fiber inertia (linear density) increased, the ability of the fiber to be a representative of the flow or to measure two-point flow statistics diminishes. Few analogies on the same lines can be drawn here: the denser fiber sees less changes in the flow when $De$ increases, whereas the polymeric fluid flow influences the flapping behaviour of the neutrally-bouyant fibers.

\subsubsection{Curvature of the fibers}
The curvature $\kappa$ exhibited is another characteristic feature representative of the fiber whose relevance stems from the fact that it is a measure of the flexibility of the body. For example, it was shown experimentally that the extent of flexibility of a filament changes its transport properties in a cellular flow \citep{wandersman2010buckled}, and thus tracking the effects of the relevant control parameters on this observable can possibly be a pathway to identify the overall system dynamics. 

\begin{figure}
 \begin{subfigure}[b]{0.5\textwidth}
  \includegraphics[width=1\textwidth,trim={1cm 0 40 15},clip]{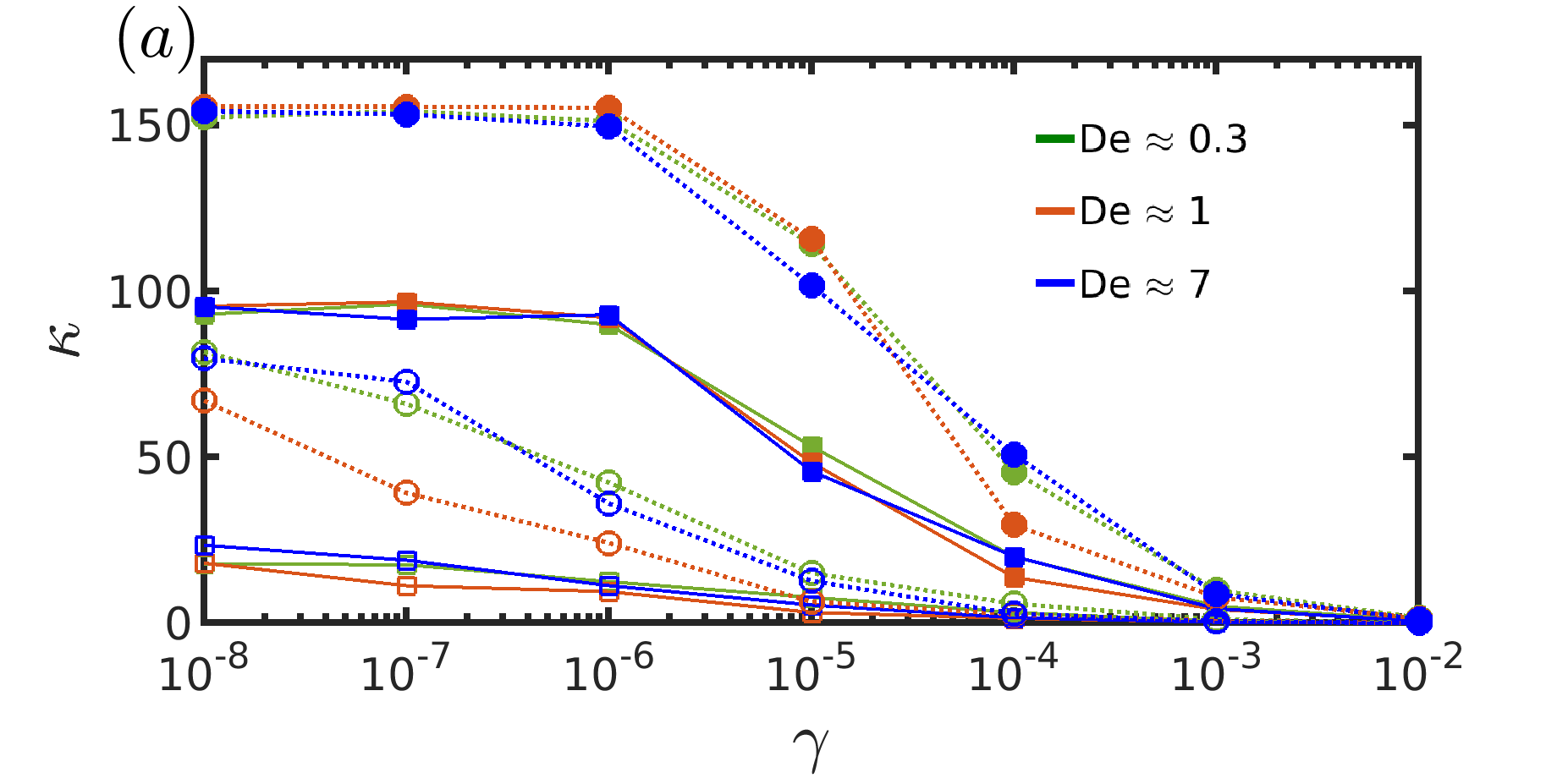}
%  \caption{}
 \end{subfigure}
 \begin{subfigure}[b]{0.5\textwidth}   
 \includegraphics[width=1\textwidth,trim={0cm 0 30 15},clip]{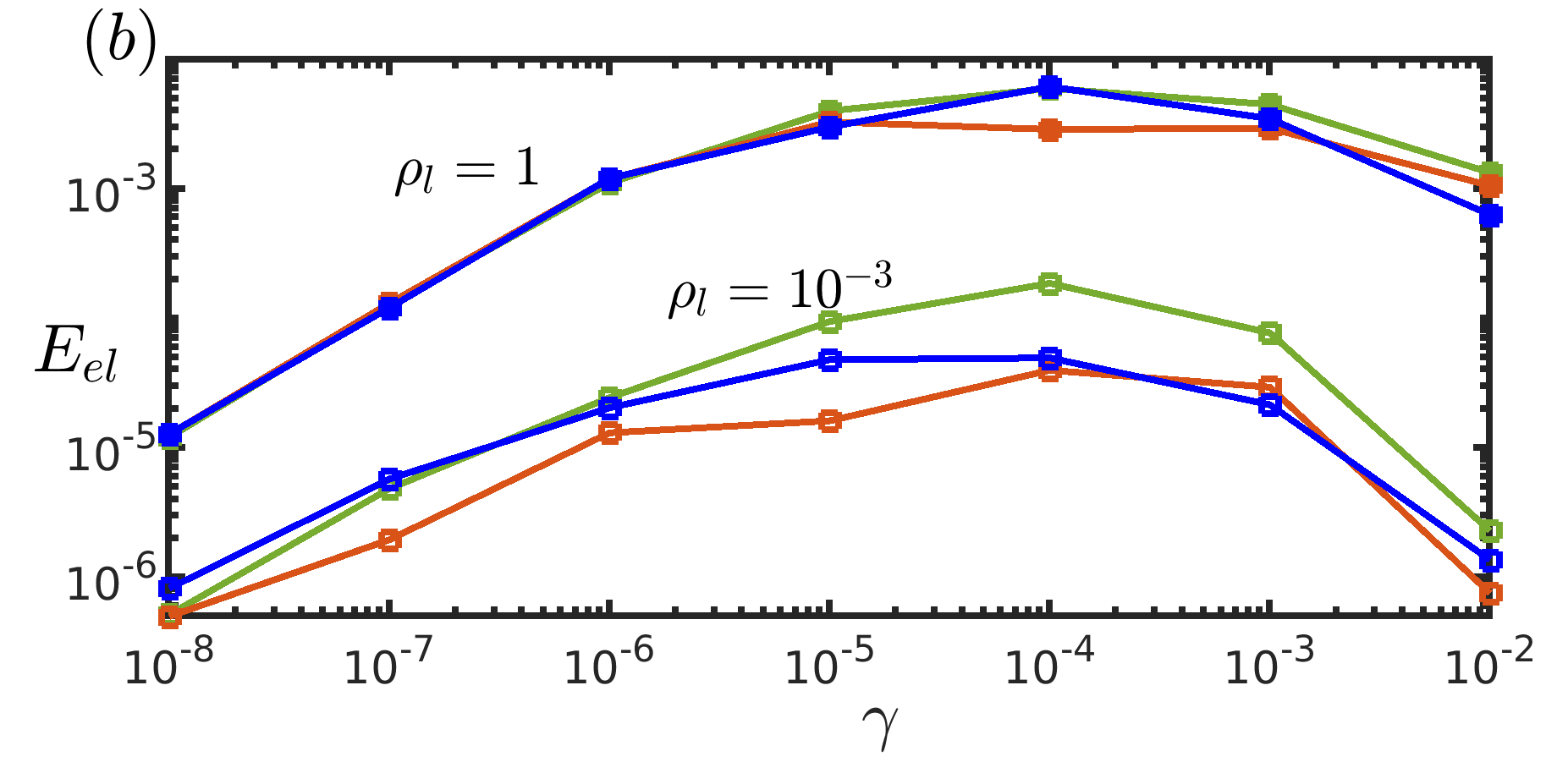}% Here is how to import EPS art
 %\caption{}
 \end{subfigure} \caption{\label{fig:epsart} (a) The curvature and (b) elastic energy stored by the neutrally buoyant and denser fibers at various $De$. \textcolor{black}{Open and closed symbols are used to distinguish neutrally bouyant and denser cases, respectively, while in (a) we use dotted and solid lines to distinguish the maximum and mean curvatures.}}
 \label{el}
\end{figure}

$\kappa$ is defined as $\sqrt{(x''(s)^2)+(y''(s)^2)+(z''(s)^2)}$, where $(x, y, z)$ are the Lagrangian coordinates and $''$ indicates the second derivative with respect to $s$.
\textcolor{black}{The average of the maximum curvature of each fiber calculated over different Lagrangian points over a few snapshots of time is plotted (dotted lines) in Fig~\ref{el}a, for both the neutrally buoyant (open circles) and the denser case (closed circles). Also, the average of the mean curvature is represented (solid lines) for neutrally bouyant (open squares) and denser case (closed squares)}. The denser-than-fluid fibers show a higher curvature compared to the neutrally bouyant case, a consequence of the former being more inertial. \textcolor{black}{This can be interpreted by balancing inertial ${\rho_{l}} \ddot{\mathbf{X}}$ and bending forces $\gamma \partial_{s}^{4} \mathbf{X}$; being  $\partial_{s}^{2} \mathbf{X}$ the curvature, it can be seen that as $\rho_{l}$ increases, the curvature can indeed increase for the same $\gamma$.}
The curvature $\kappa$ decreases with increasing stiffness $\gamma$, being almost constant with high values for flexible fibers, transitioning into lower values for highly stiff fibers through an intermediate regime with interim values. The variation of the curvature with $De$ shows a non-monotonic trend at the intermediate $De$ in the neutrally bouyant scenario. %The flapping configurations of the fiber\citep{connell2007flapping},
The elastic energy stored by the fibers defined as $E_{el}$ = $\int_0^c \frac{1}{2} \gamma \kappa ^2 ds $ is plotted in Fig \ref{el}b. The denser fibers have higher elastic energy, consistent with their overall larger curvature, but seem to be unaffected by $De$, whereas an evident drop is exhibited by the neutrally-bouyant fibers as $De$ changes from 0.3 (low) to 1 (interim), which recovers when $De$ is increased further.

\begin{figure}
\begin{subfigure}[b]{0.5\textwidth}
\hspace{-0.5 cm}
\includegraphics[width=1.1\textwidth,trim={1cm 8 8 8},clip]{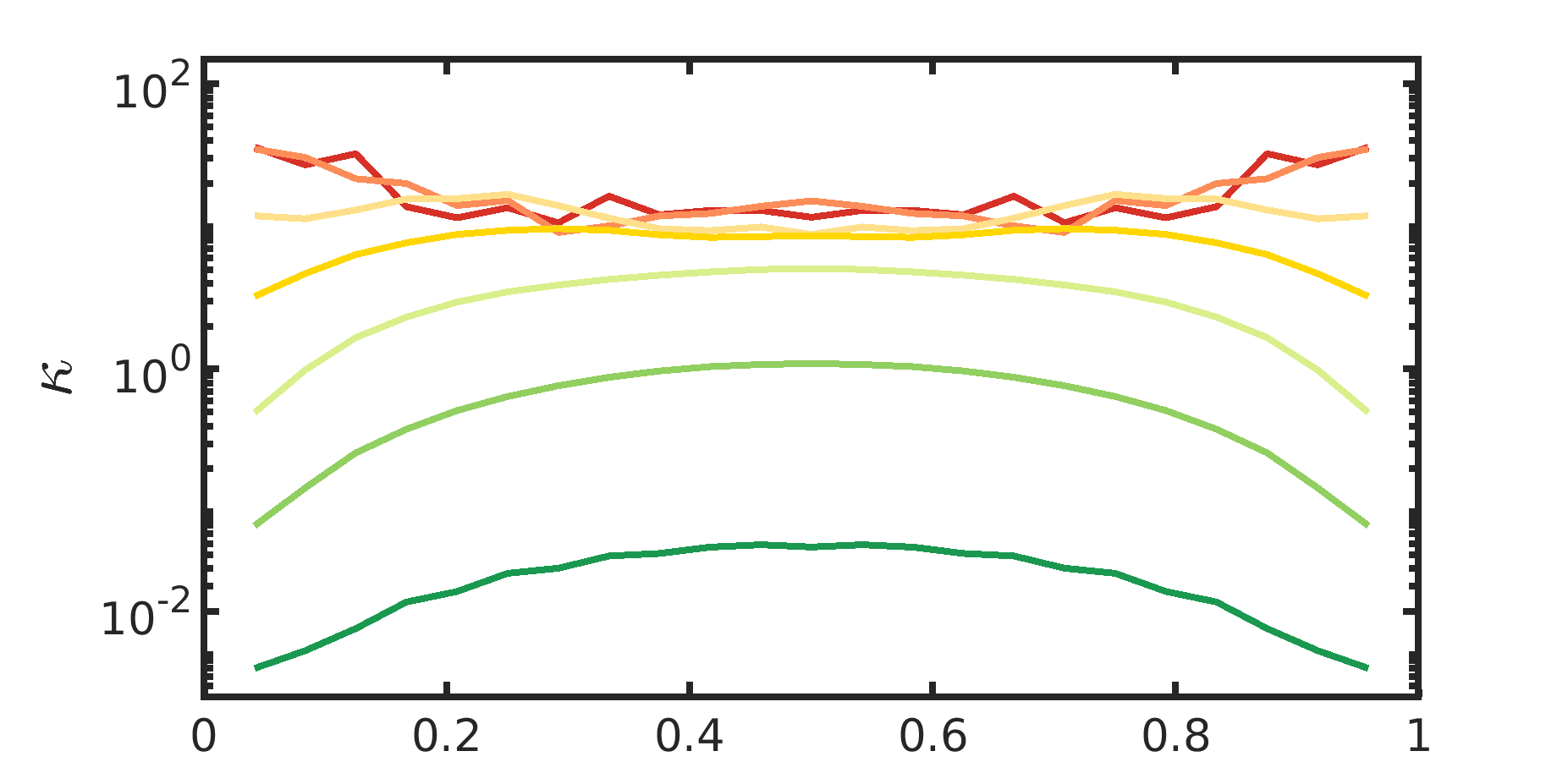}% Here is how to import EPS art
\caption{}
\end{subfigure}
\begin{subfigure}[b]{0.5\textwidth}
\hspace{-0.5 cm}
\includegraphics[width=1.1\textwidth,trim={1cm 8 8 8},clip]{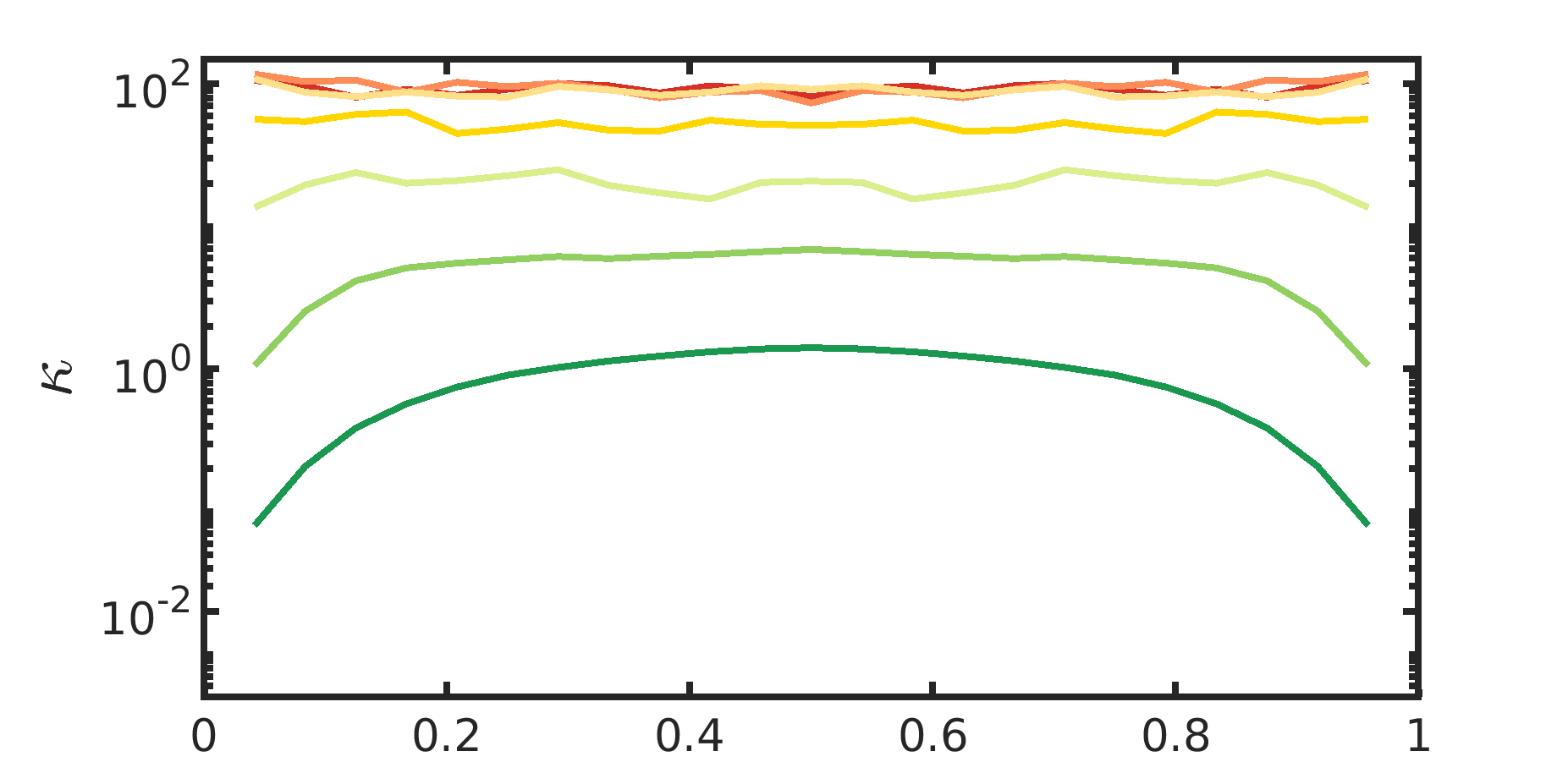}% Here is how to import EPS art
\caption{}
\end{subfigure}
\begin{subfigure}[b]
{0.5\textwidth}
\hspace{-0.5 cm}
\includegraphics[width=1.1\textwidth,trim={1cm 8 8 8},clip]{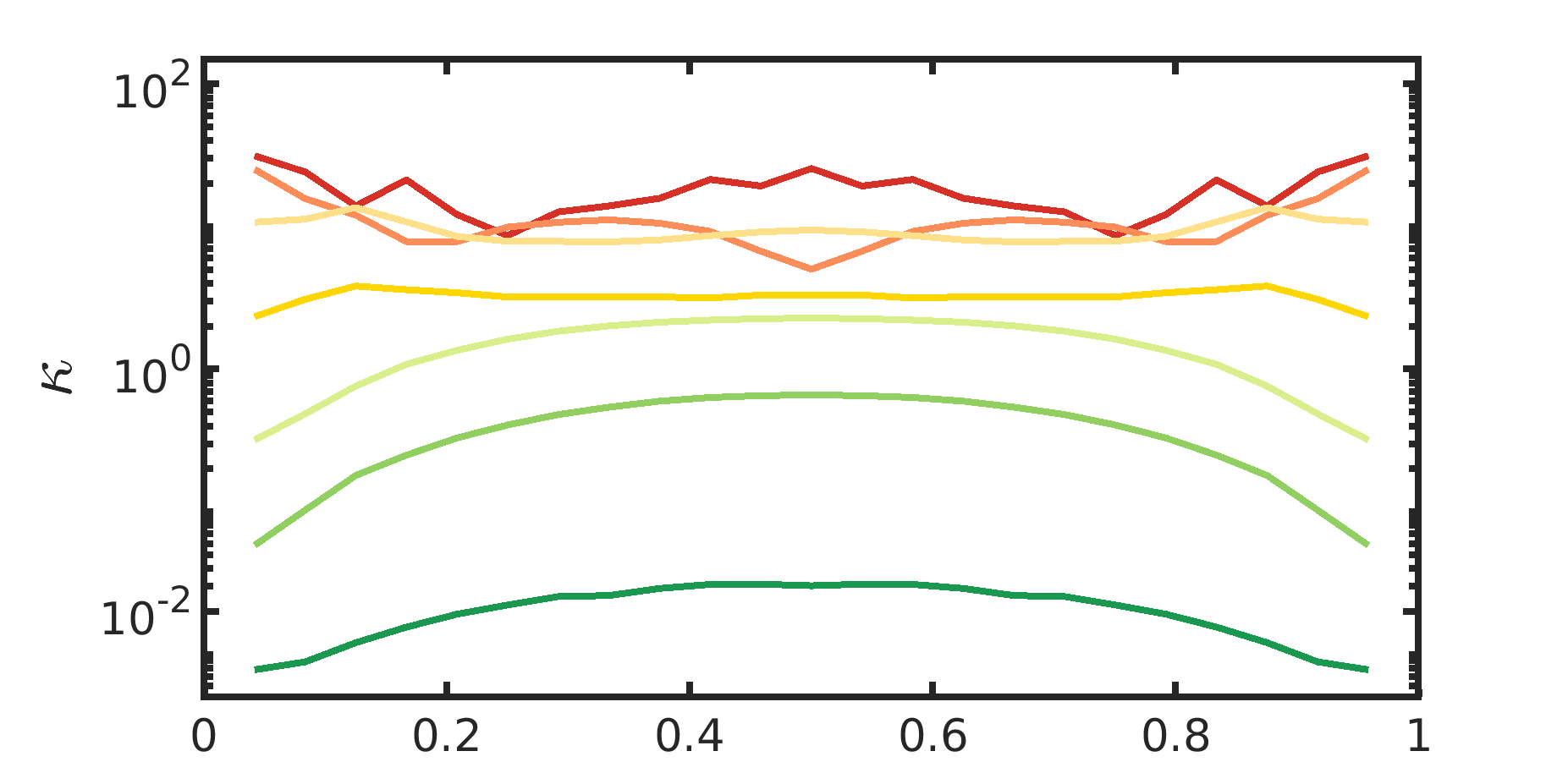}% Here is how to import EPS art
\caption{}
\end{subfigure}
\begin{subfigure}[b]{0.5\textwidth}
\hspace{-0.5 cm}
\includegraphics[width=1.1\textwidth,trim={1cm 8 8 8},clip]{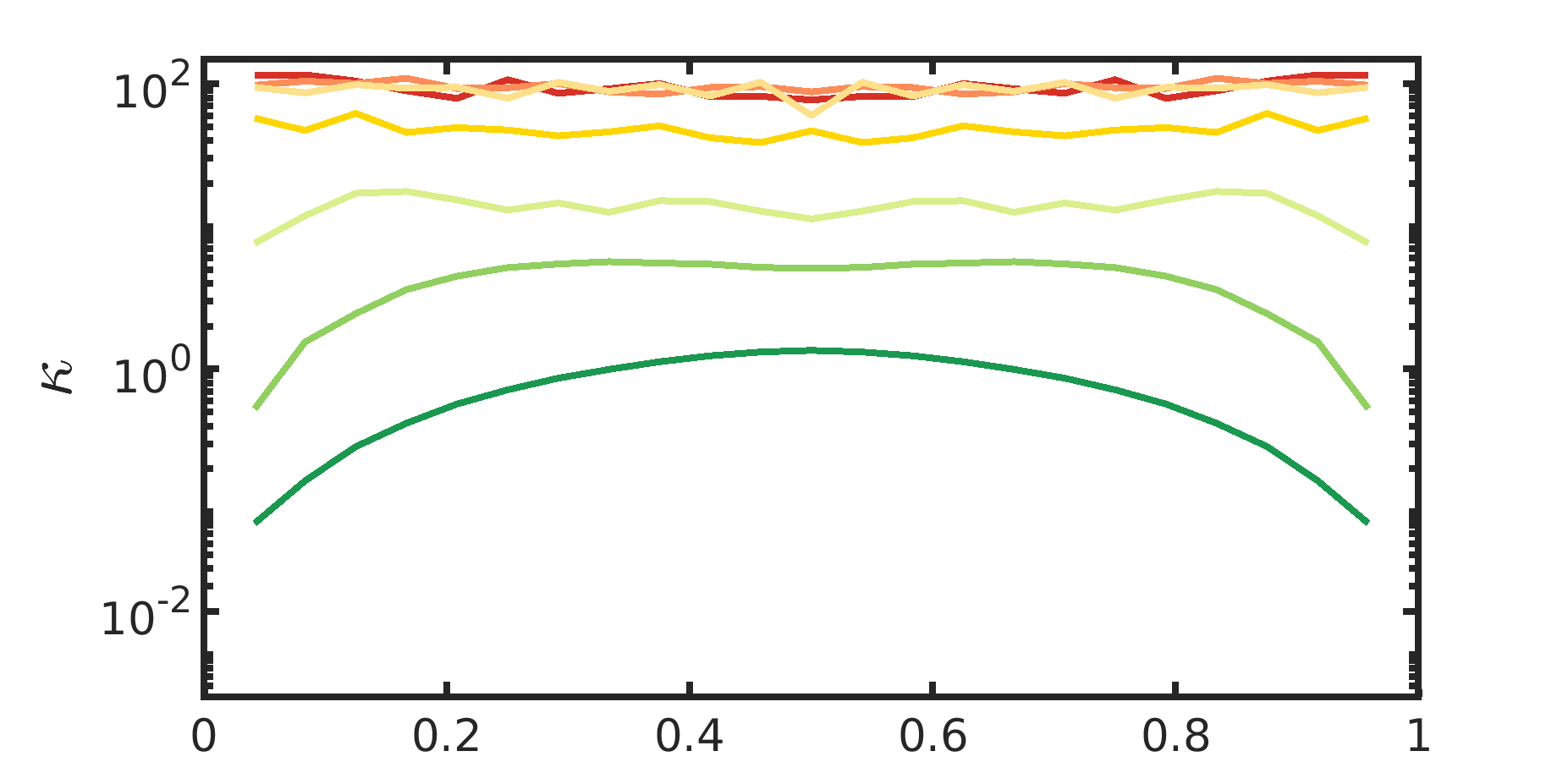}% Here is how to import EPS art
\caption{}
\end{subfigure}
\begin{subfigure}[b]{0.5\textwidth}
\hspace{-0.5 cm}
\includegraphics[width=1.1\textwidth,trim={1cm 3 8 8},clip]{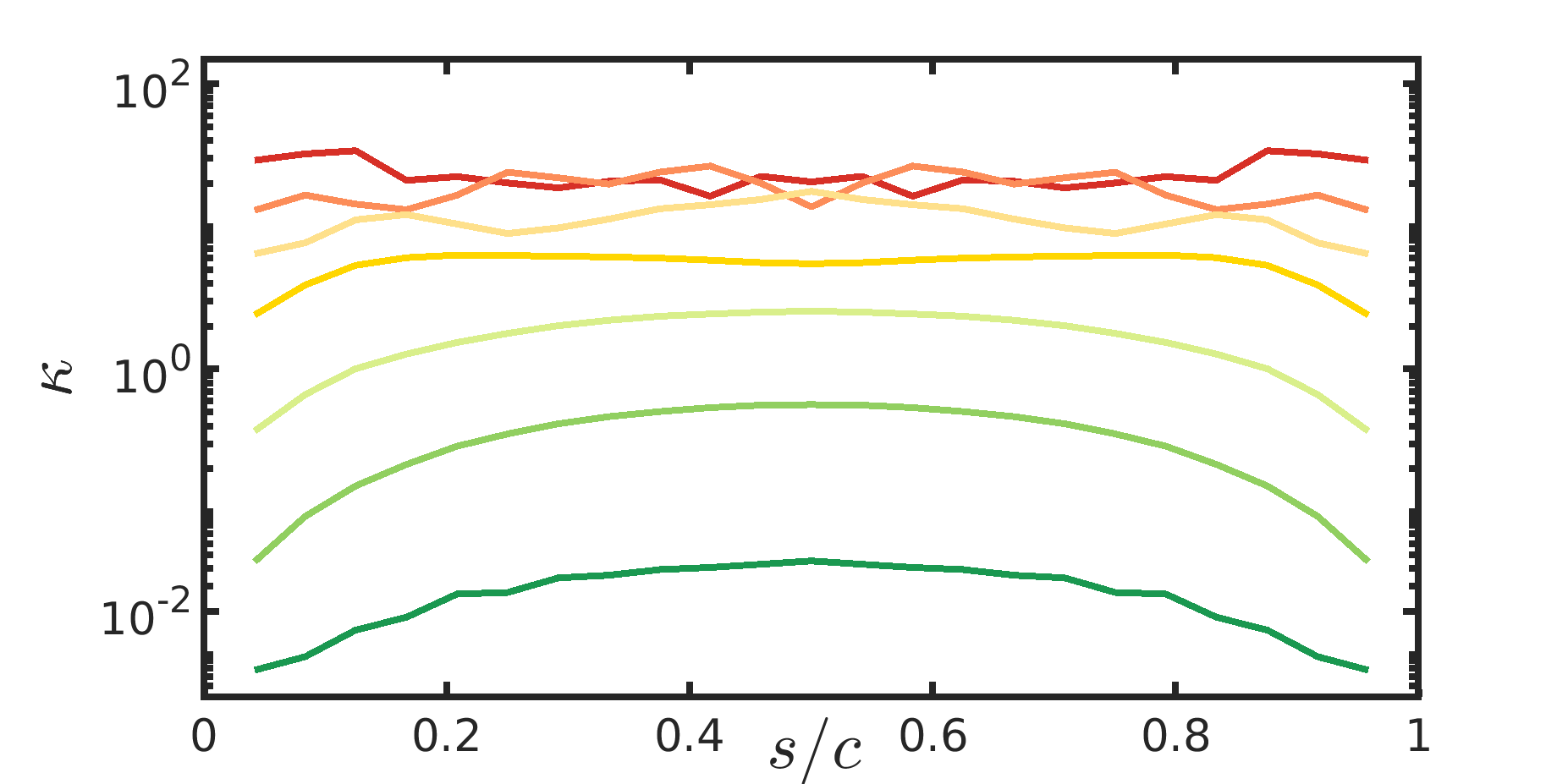}% Here is how to import EPS art
\caption{}
\end{subfigure}
\begin{subfigure}[b]{0.5\textwidth}
\hspace{-0.5 cm}
\includegraphics[width=1.1\textwidth,trim={1cm 3 8 8},clip]{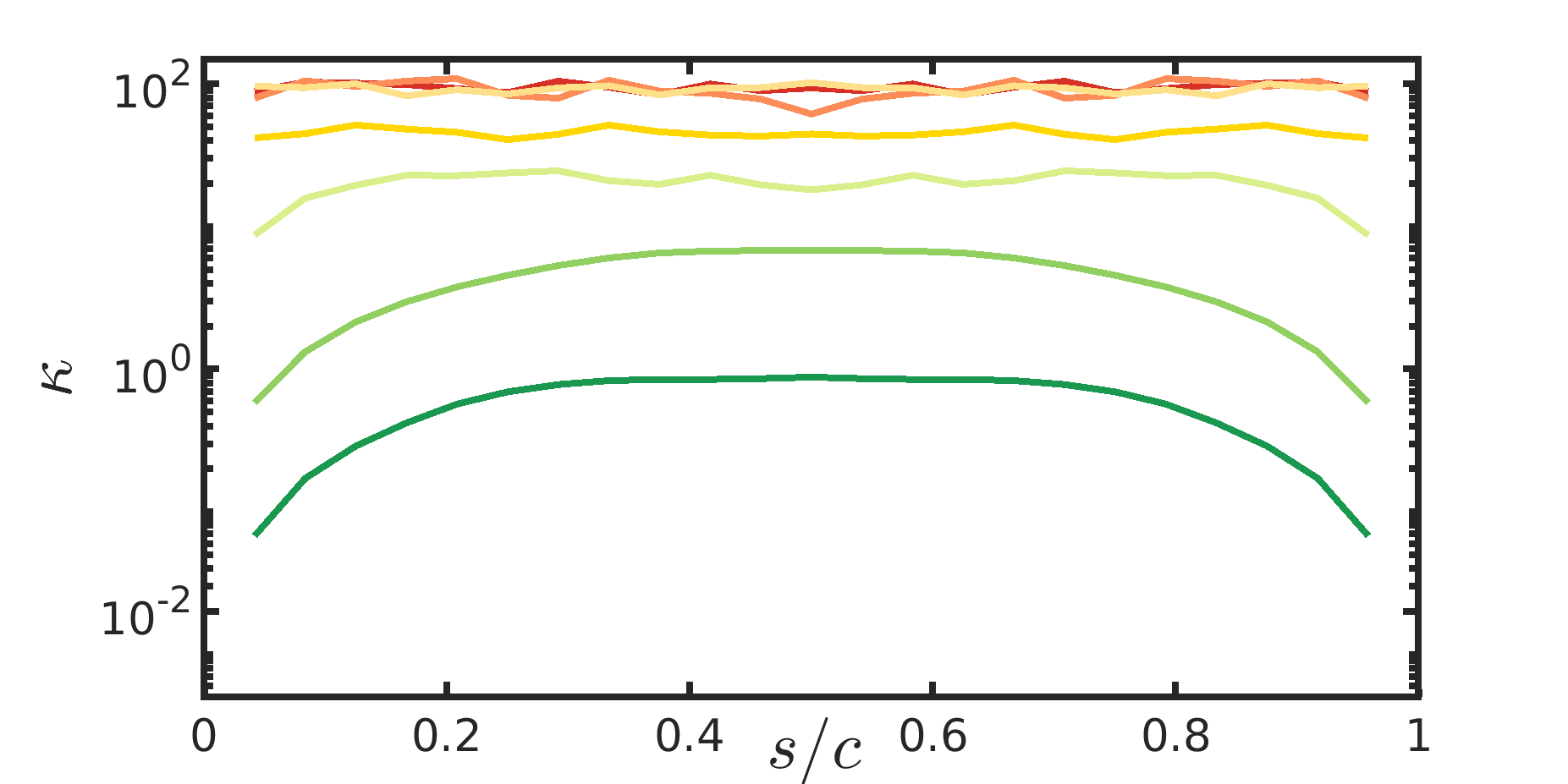}% Here is how to import EPS art
\caption{}
\end{subfigure}
\begin{subfigure}[b]{0.8\textwidth}
\hspace{1 cm}
\includegraphics[width=1\textwidth,trim={0cm 0 0 0},clip]{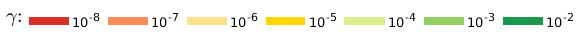}% Here is how to import EPS art
\end{subfigure}
\caption{The curvature $\kappa$ plotted as a function of the normalised length $s/c$, along the fibre length at (a, b) $De$ $\approx$ 0.3, (c, d) $1$, and (e, f) $7$. Left and right panels show the neutrally-bouyant and denser cases.}
\label{fig:curvs}
\end{figure}

To better perceive how the fiber shape deforms, we further probe into the fiber curvature by plotting in Fig \ref{fig:curvs}a temporally averaged $\kappa$ of each fiber as a function of its normalised length $s/c$. The left and right panels respectively show the neutrally bouyant and denser cases with the top (Fig \ref{fig:curvs}a, b), middle (Fig \ref{fig:curvs}c, d), and bottom panels (Fig \ref{fig:curvs}e, f) representing $De$ $\approx$ 0.3, 1, and 7, respectively. Clearly, as rigidity increases, $\kappa$ decreases for all cases. For the highest $\gamma$, the neutrally-bouyant fibers are almost in an unbent configuration in comparison with the denser ones at the highest $\gamma$ (compare the abscissa of Fig \ref{fig:curvs}a, b.), which exhibits a unimodal shape. The denser fiber further becomes bimodal and multimodal from thereon as $\gamma$ decreases, eventually becoming a highly flexible and deformable body. This happens also for the neutrally bouyant fibers, but at much lower $\gamma$: indeed, for the same rigidity, the deformation and magnitudes of the neutrally bouyant fibers are lower compared to the denser ones.
% \begin{figure}
% %\centering
% \begin{subfigure}[b]{0.5\textwidth}
% \centering \includegraphics[width=1\textwidth,trim={1cm 0 6 0},clip]{figures/curvavgDlowiso.eps}% Here is how to import EPS art
% \caption{\label{fig:epsart}neutrally-bouyant}
% \end{subfigure}
% \begin{subfigure}[b]{0.5\textwidth}
% \centering \includegraphics[width=1\textwidth,trim={1cm 0 6 0},clip]{figures/curvavgDlowdense.eps}% Here is how to import EPS art
% \caption{\label{fig:epsart}denser}
% \end{subfigure}
% \begin{subfigure}[b]
% {0.5\textwidth}
% \centering
% \includegraphics[width=1\textwidth,trim={1cm 0 6 0},clip]{figures/curvavgDmediso.eps}% Here is how to import EPS art
% \caption{\label{fig:epsart}neutrally-bouyant}
% \end{subfigure}
% \begin{subfigure}[b]{0.5\textwidth}
% \includegraphics[width=1\textwidth,trim={1cm 0 6 0},clip]{figures/curvavgDmeddense.eps}% Here is how to import EPS art
% \caption{\label{fig:epsart}denser}
% \end{subfigure}
% \end{figure}

%One of the important characteristic features often of interest while studying fibers is when and how the transition to buckling happens. 
Fig \ref{fig:curvs} a, b ascertains that for the same stiffness, fibers of higher linear density buckle to a higher extent in comparison to the neutrally-bouyant ones (see the green colors). In other words, a dynamical transition to buckling from an unbent configuration is easily initiated when the fibers are more inertial. When investigating if viscoelasticity further hinders this behavior, we observe that the deformations of the neutrally-bouyant fibers (Fig \ref{fig:curvs} a, c, e) are slightly affected by viscoelasticity, while the denser fibers are insensitive to variations in fluid elasticity (Fig \ref{fig:curvs} b, d, f).
It is known that capsules in an Oldroyd-B shear flow experience monotonically decreasing or increasing deformations depending on the level of elasticity, and that the 3D flapping of a flag is hindered by viscoelasticity \citep{ma2020immersed}. \cite{fu2008beating} defined a bending scale for filament-like swimmers (higher for stiff filaments and smaller for flexible filaments deforming due to fluid forces), and reported that viscoelasticity increased this bending scale. This was correlated to the beating patterns of swimmers changing from traveling waves to standing waves as Deborah number increased. All these studies were conducted at lower Reynolds numbers, where the fluid turbulence was not relevant, whereas the present study extends this understanding to turbulent scenarios in a broader parametric setting. %density of filament less than flow, re based on length of fil<5. \cite{musielak2009nutrient}-this ref is contradictin
Scrutinizing the effects of viscoelasticity on the filament deformations along similar lines as the above-discussed studies leads one to conclude that, increased polymer stretching can quantitatively influence the fiber curvature, but does not impact them qualitatively.

\subsubsection{Alignment}
\begin{figure}
%\centering
\begin{subfigure}[b]{0.8\textwidth}
\hspace{5.5cm}
\includegraphics[width=0.3\textwidth,trim={0cm 0 0 0},clip]{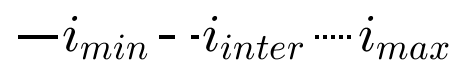}% Here is how to import EPS art
\end{subfigure}
\begin{subfigure}[b]{0.5\textwidth}
\centering %\includegraphics[width=1\textwidth,trim={1cm 0 6 0},clip]{figures/aliignDlowiso_Ap23.eps}% 
%Here is how to import EPS art
\includegraphics[width=1.1\textwidth,trim={1cm 0 6 0},clip]{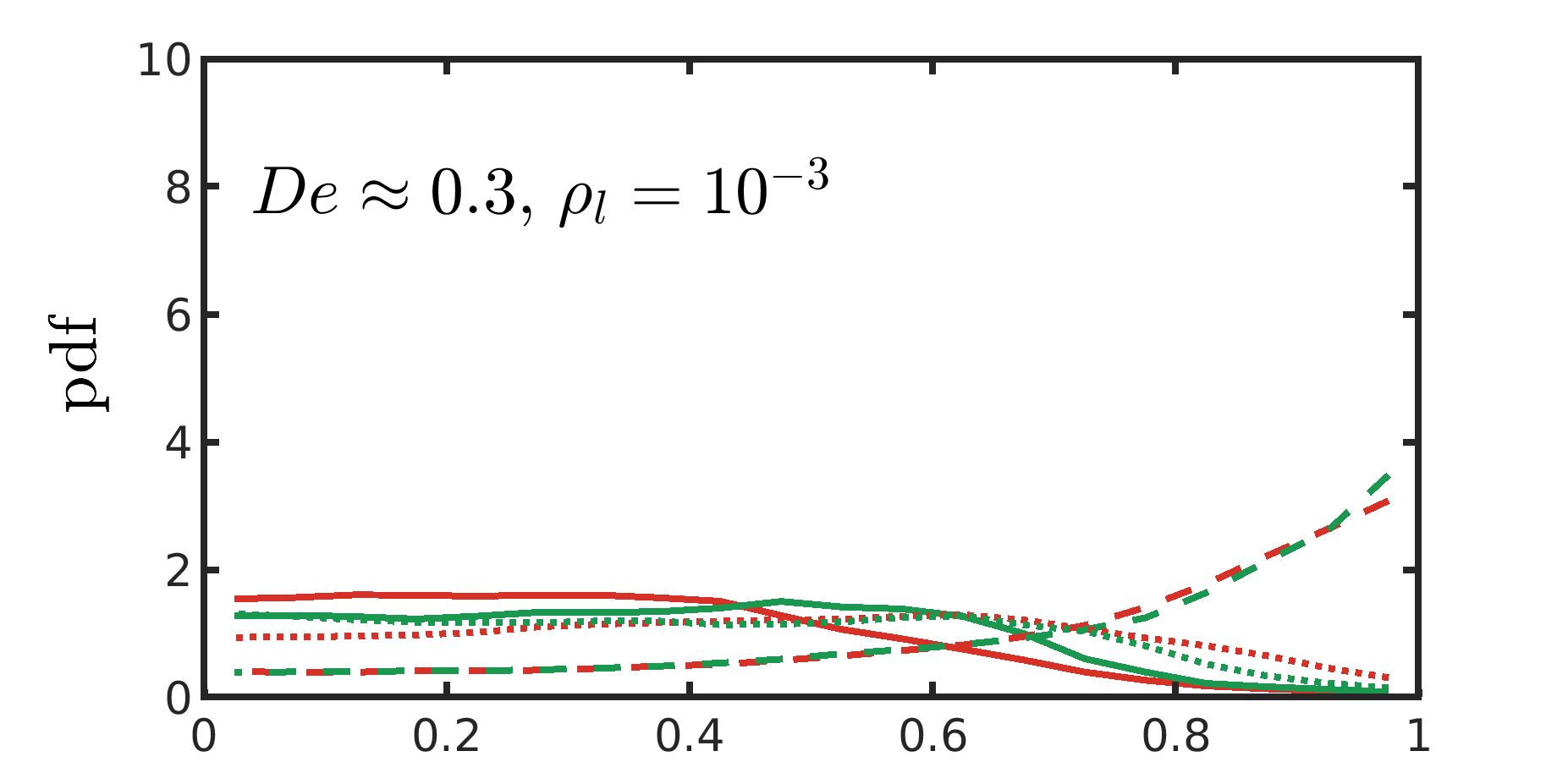}% 
\caption{}
\end{subfigure}
\begin{subfigure}[b]{0.5\textwidth}
\centering %\includegraphics[width=1\textwidth,trim={1cm 0 6 0},clip]{figures/aliignTraceDlowiso_Ap23.eps}
\includegraphics[width=1.1\textwidth,trim={1cm 0 6 0},clip]{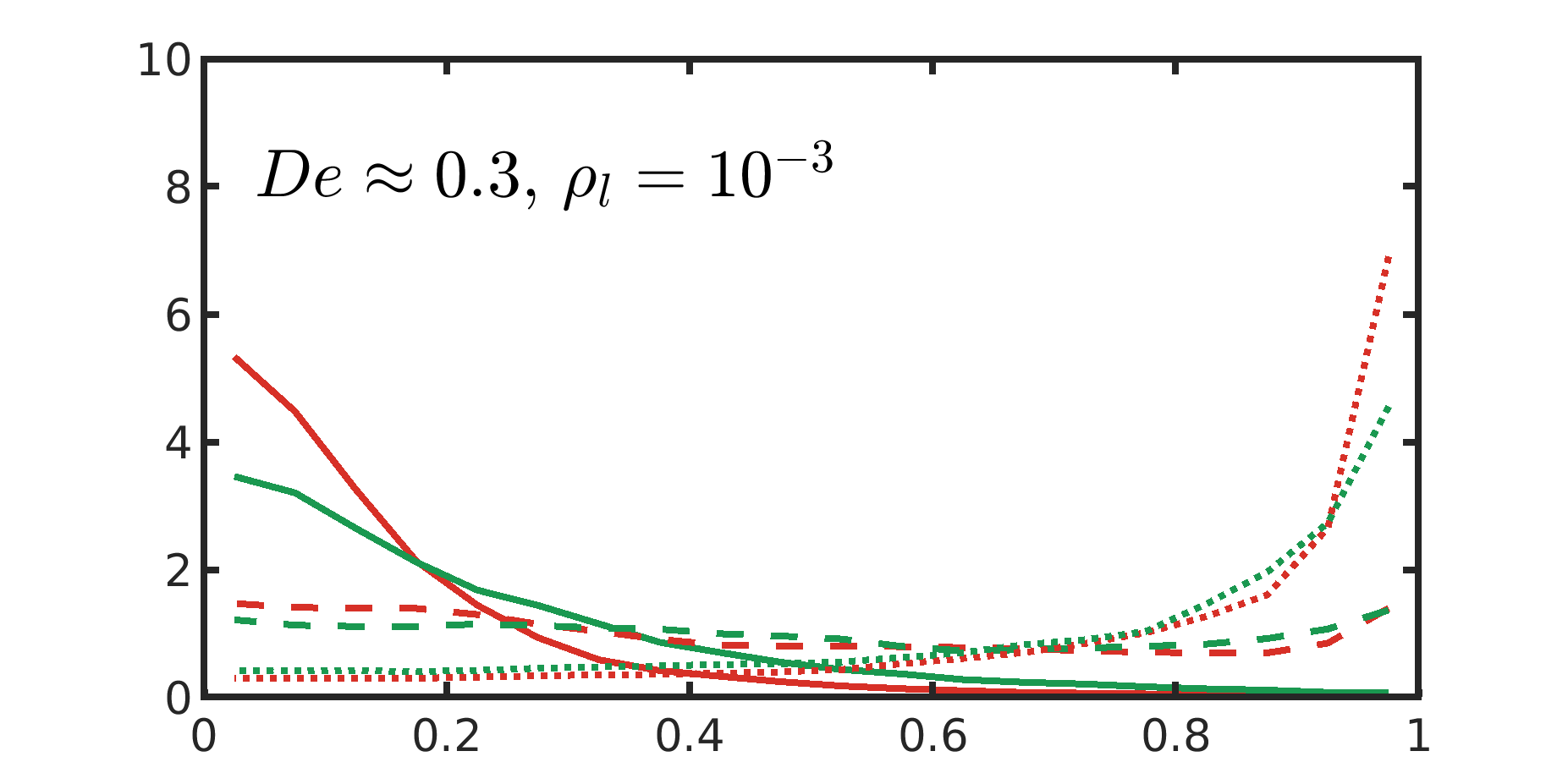}
% Here is how to import EPS art
\caption{}
\end{subfigure}
\begin{subfigure}[b]
{0.5\textwidth}
\centering
\includegraphics[width=1.1\textwidth,trim={1cm 0 6 0},clip]{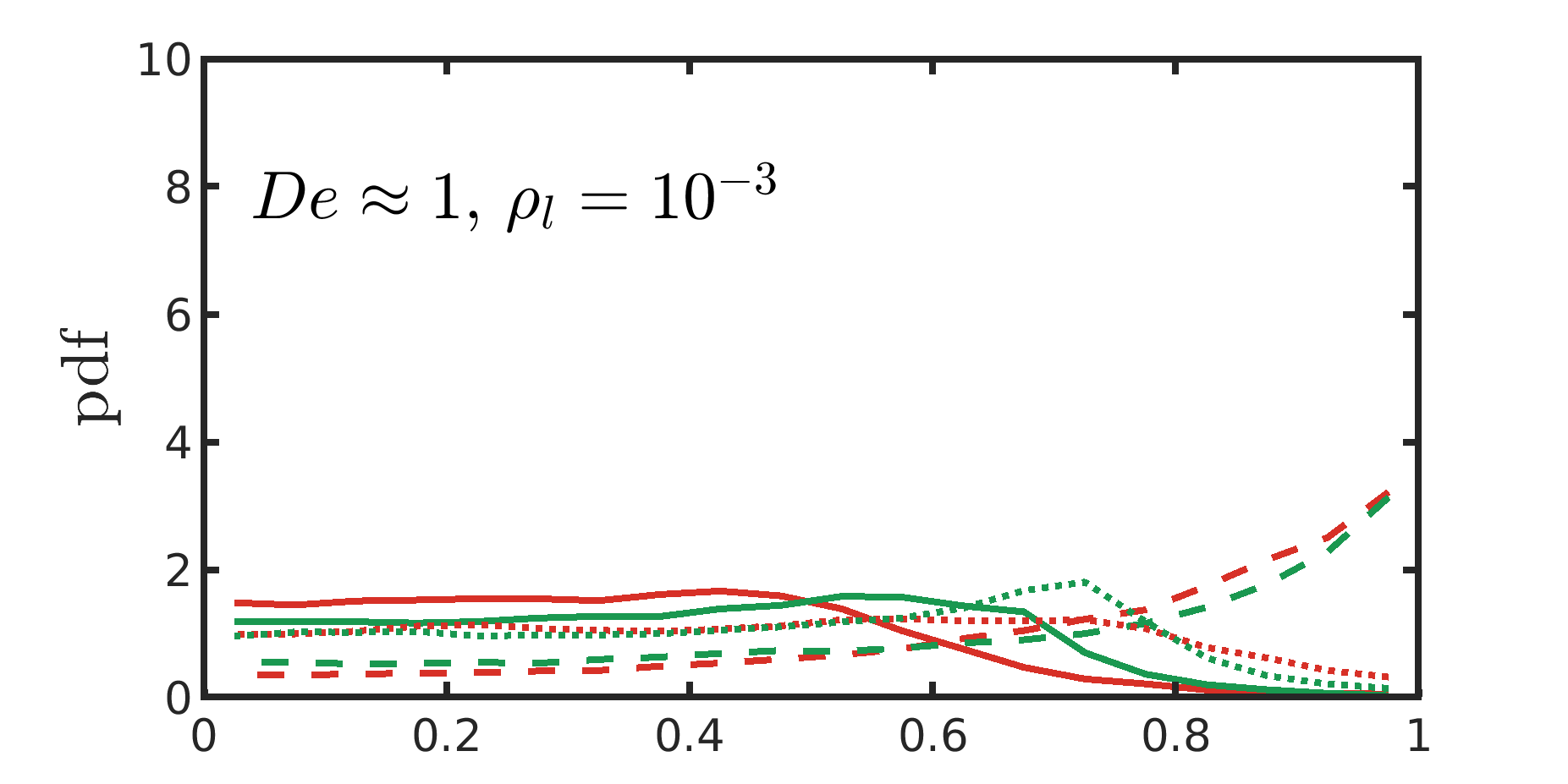}% Here is how to import EPS art
\caption{}
\end{subfigure}
\begin{subfigure}[b]{0.5\textwidth}
\includegraphics[width=1.1\textwidth,trim={1cm 0 6 0},clip]{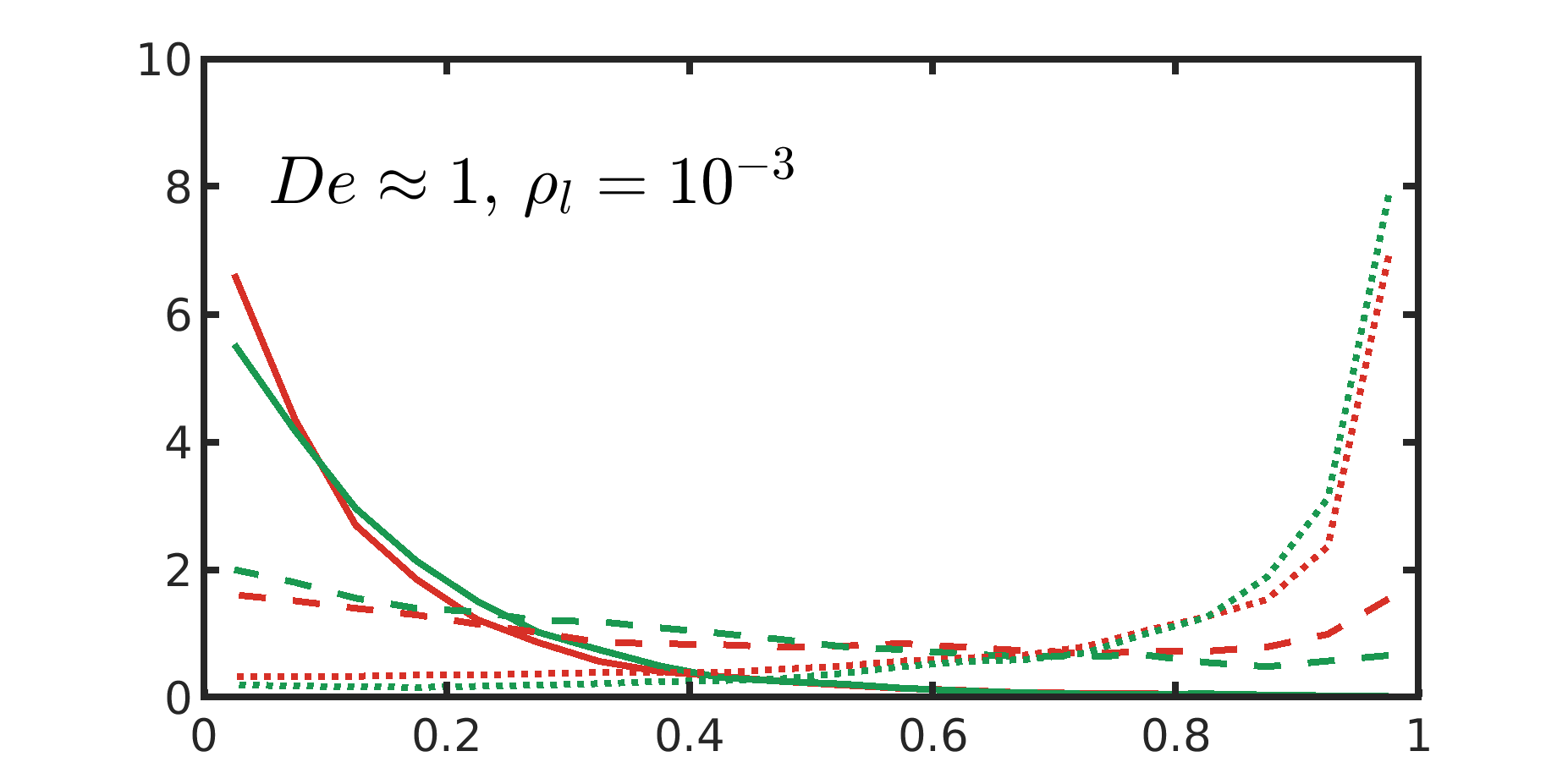}% Here is how to import EPS art
\caption{}
\end{subfigure}
\begin{subfigure}[b]{0.5\textwidth}
\includegraphics[width=1.1\textwidth,trim={1cm 0 6 0},clip]{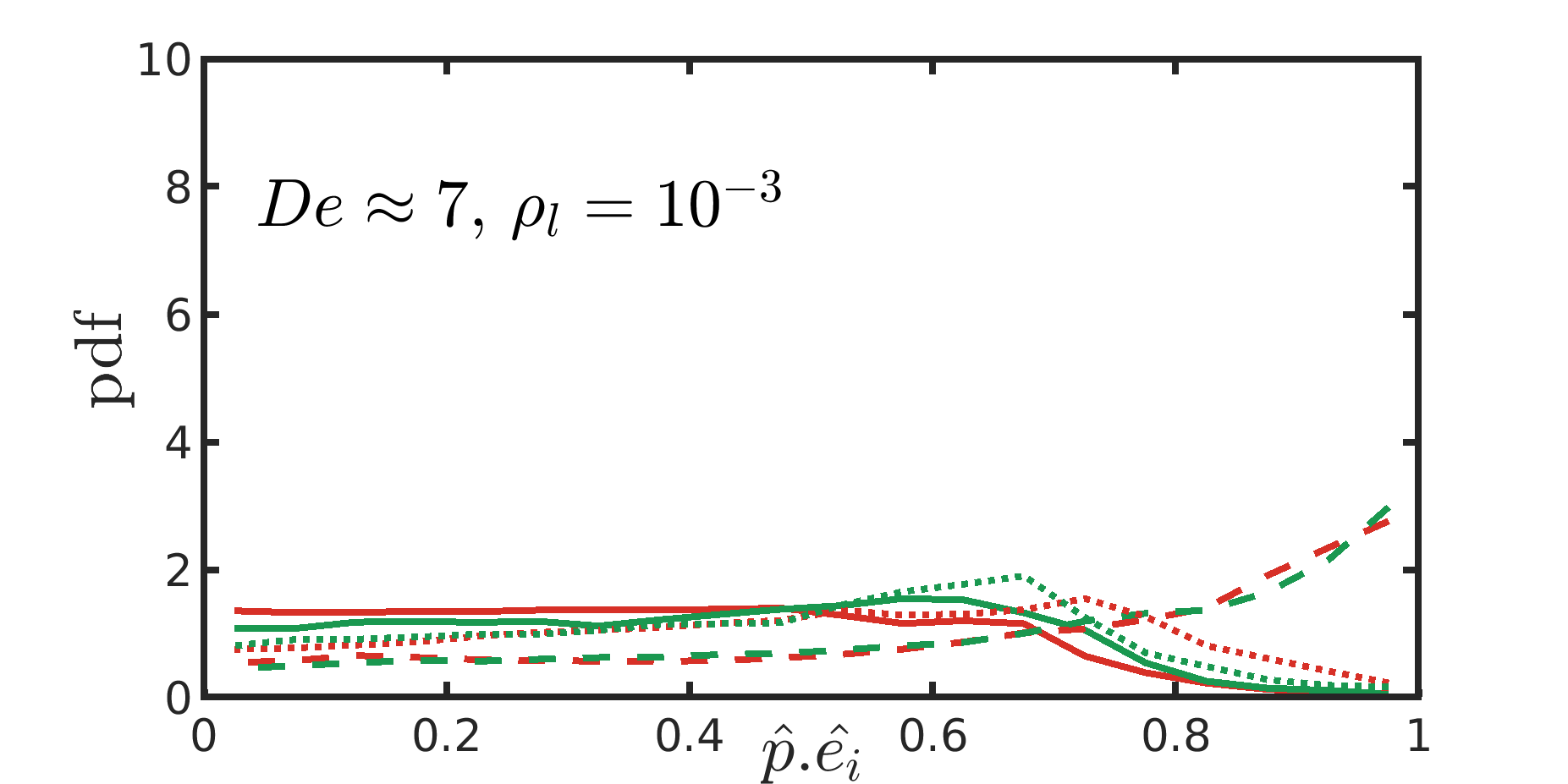}% Here is how to import EPS art
\caption{}

\end{subfigure}
\begin{subfigure}[b]{0.5\textwidth}
\includegraphics[width=1.1\textwidth,trim={1cm 0 6 0},clip]{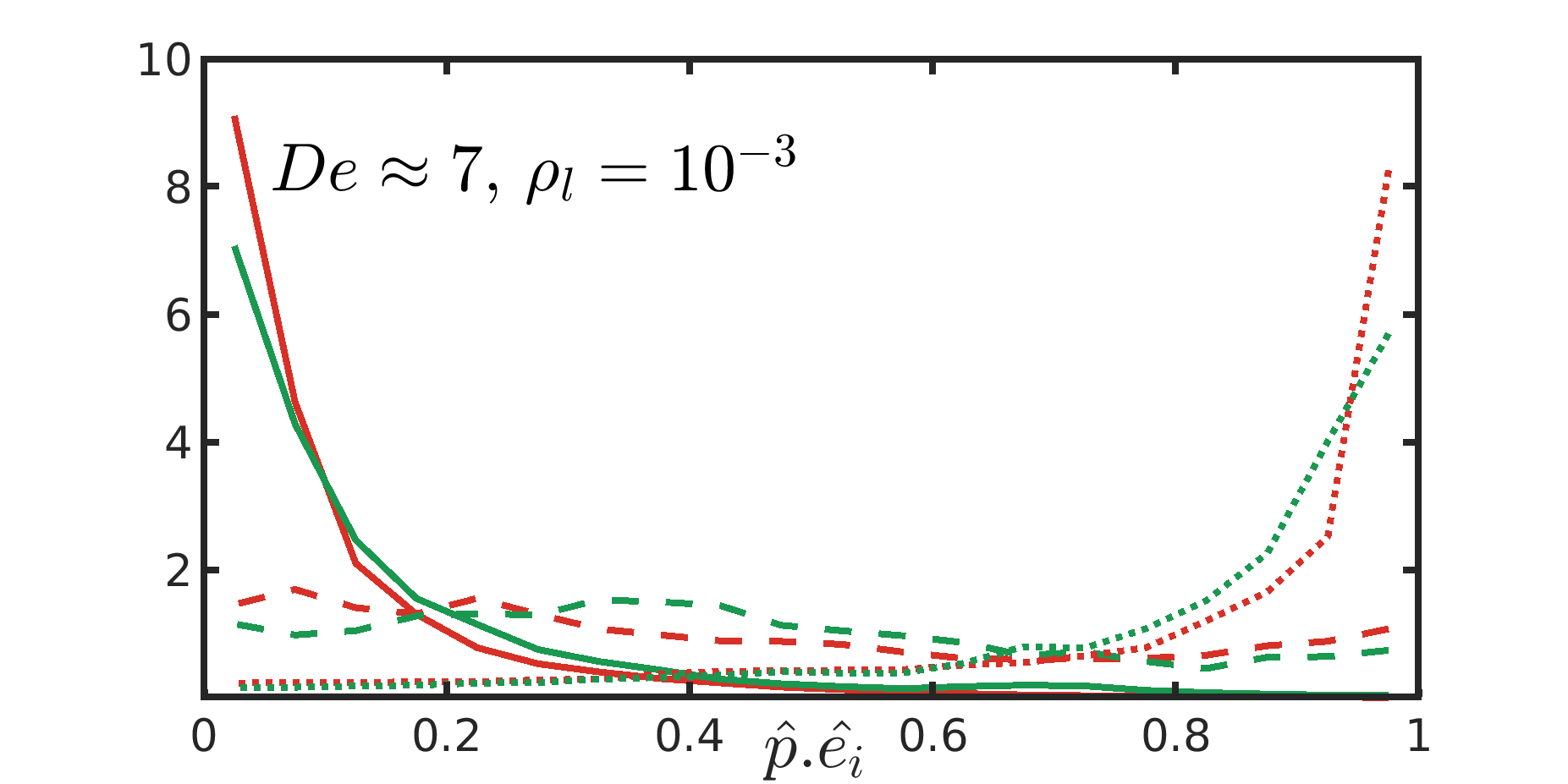}
\caption{}
\end{subfigure}
\begin{subfigure}[b]{0.8\textwidth}
\hspace{6cm}
\includegraphics[width=0.3\textwidth,trim={0cm 0 0 0},clip]{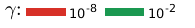}% Here is how to import EPS art
\end{subfigure}
\caption{Probability density functions of the alignment of neutrally-bouyant fibers with the principal directions of (left) the strain rate tensor and (right) the conformation tensor, at $De \approx$ (a,b) 1, (c,d) 3, and (e, f) 7. Solid, dashed and dotted lines correspond to $i_{min}$, $i_{inter}$, and $i_{max}$, respectively.}
\label{aligniso}
\end{figure}

\begin{figure}
%\centering
\begin{subfigure}[b]{0.8\textwidth}
\hspace{5.5cm}
\includegraphics[width=0.3\textwidth,trim={0cm 0 0 0},clip]{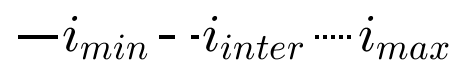}% Here is how to import EPS art
\end{subfigure}
\begin{subfigure}[b]{0.5\textwidth}
\centering \includegraphics[width=1.1\textwidth,trim={1cm 0 6 0},clip]{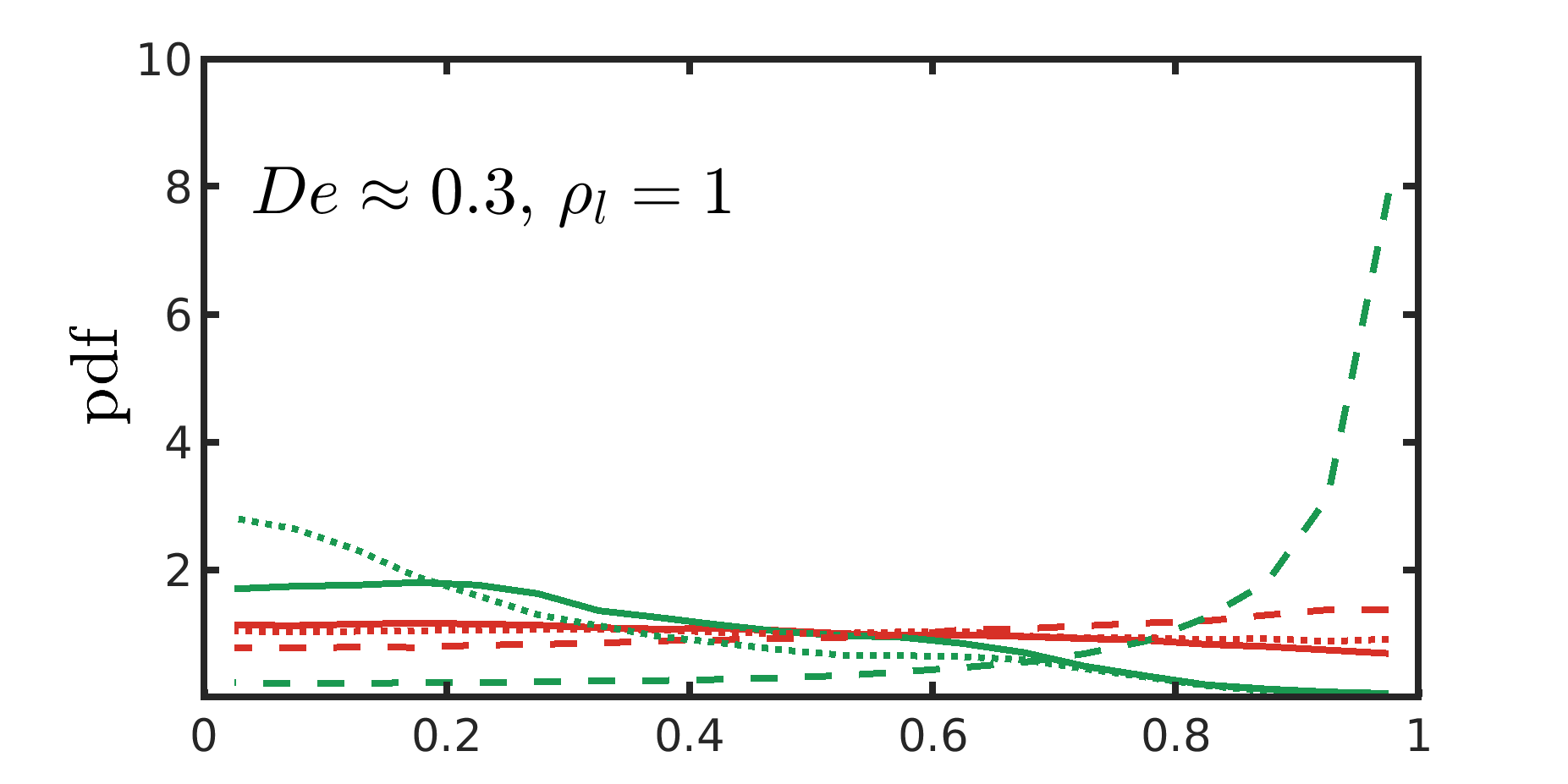}% Here is how to import EPS art
\caption{}
\end{subfigure}
\begin{subfigure}[b]{0.5\textwidth}
\centering \includegraphics[width=1.1\textwidth,trim={1cm 0 6 0},clip]{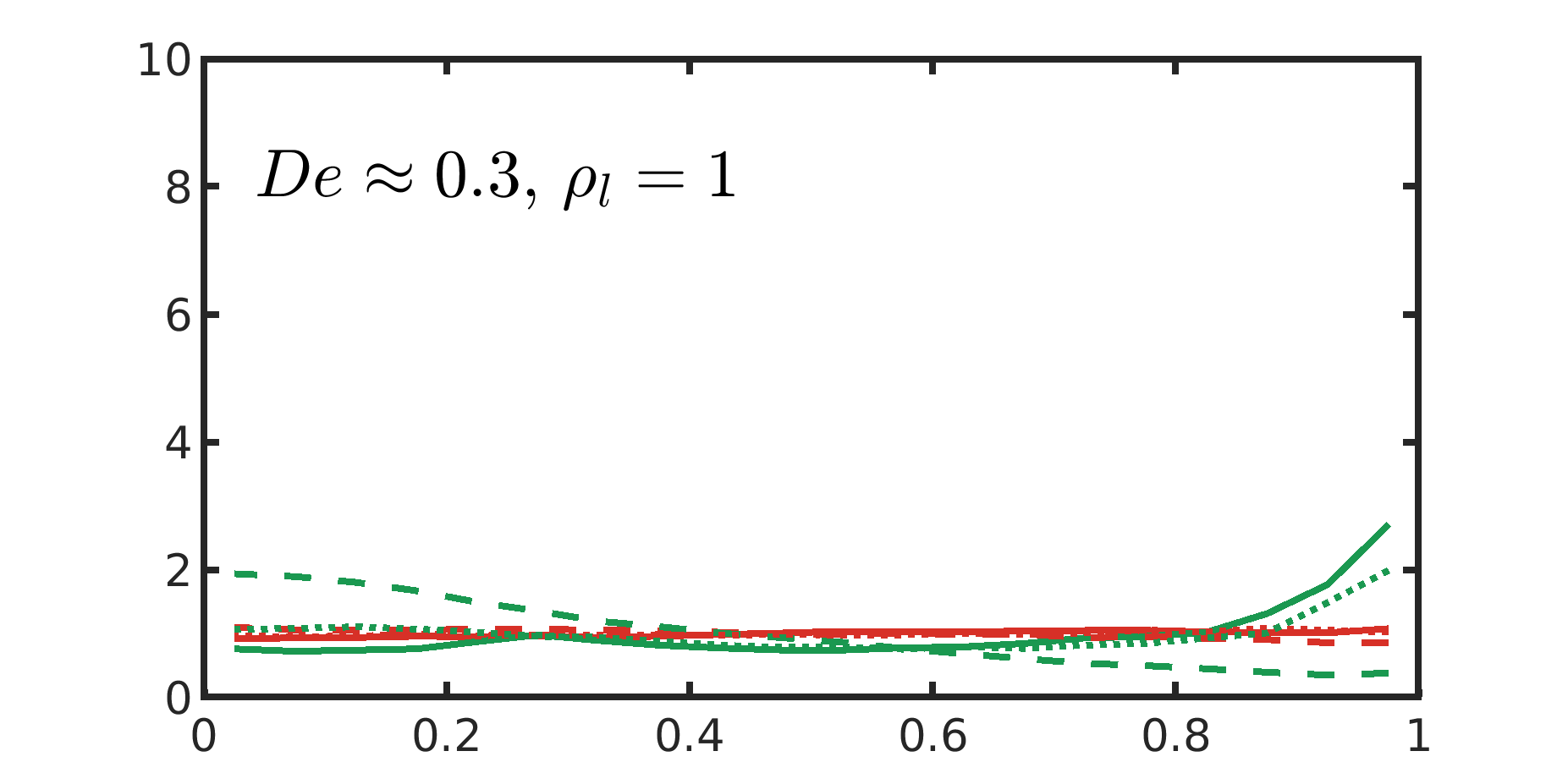}% Here is how to import EPS art
\caption{}
\end{subfigure}
\begin{subfigure}[b]
{0.5\textwidth}
\centering
\includegraphics[width=1.1\textwidth,trim={1cm 0 6 0},clip]{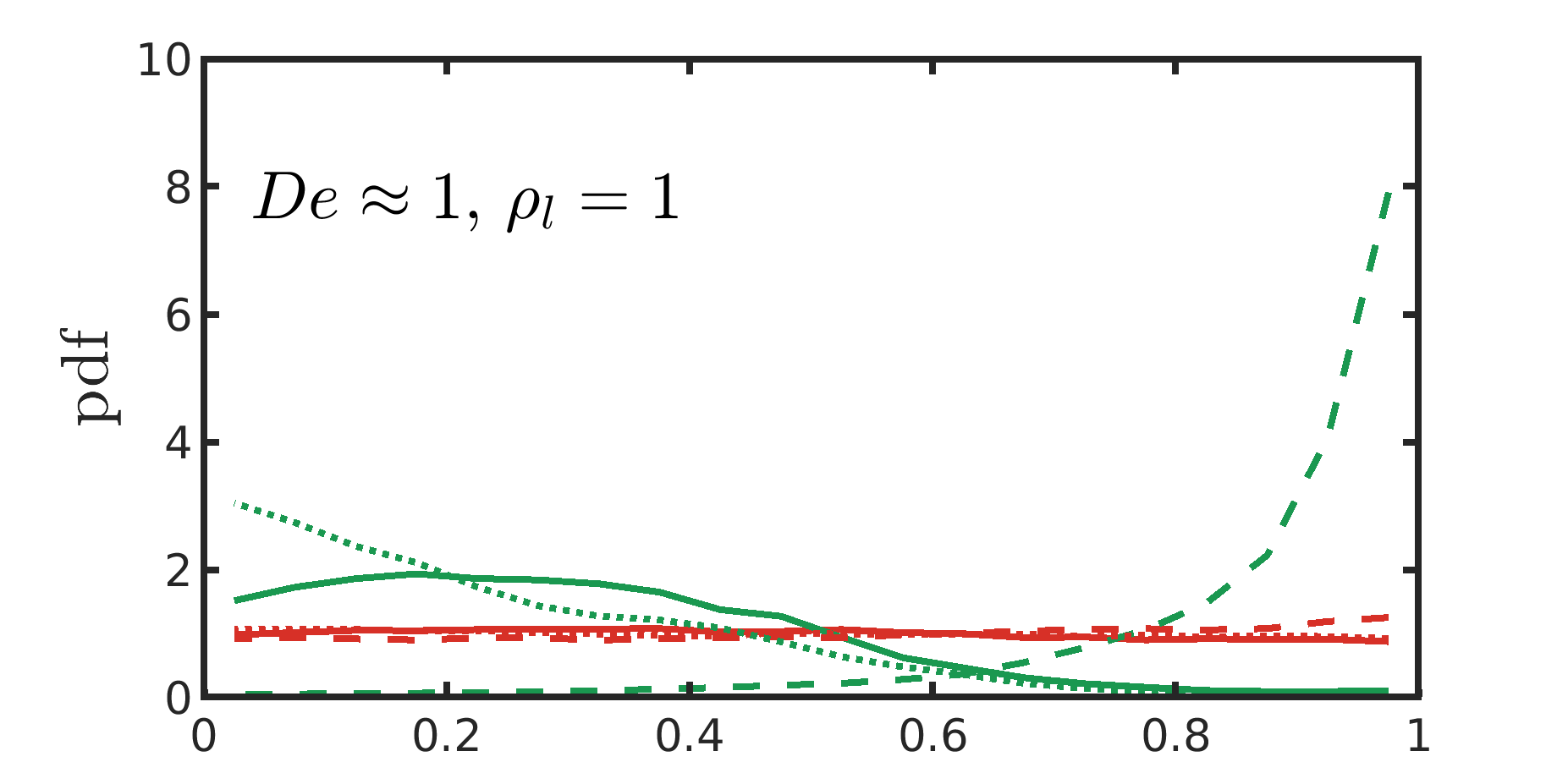}% Here is how to import EPS art
\caption{}
\end{subfigure}
\begin{subfigure}[b]{0.5\textwidth}
\includegraphics[width=1.1\textwidth,trim={1cm 0 6 0},clip]{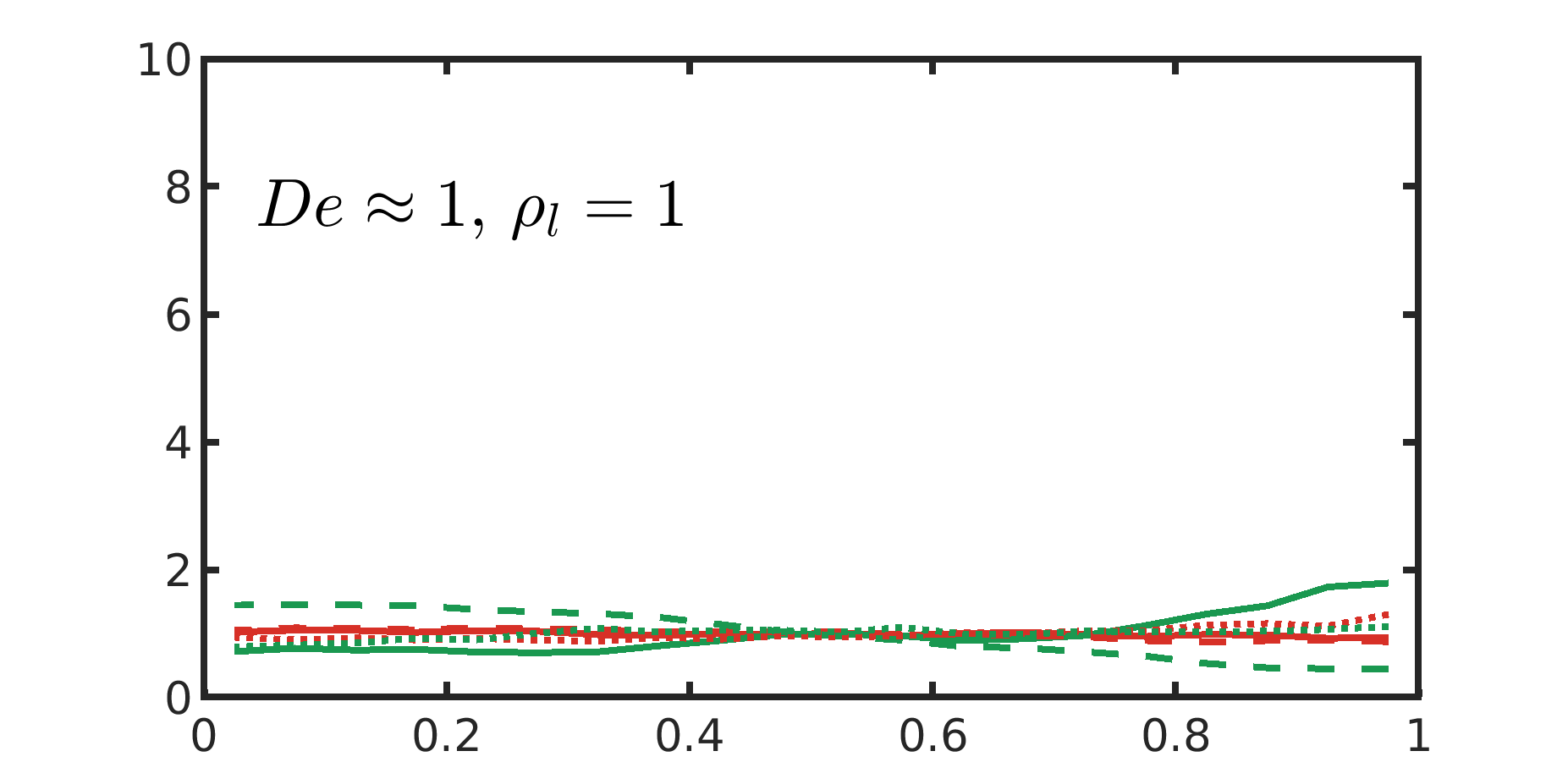}% Here is how to import EPS art
\caption{}
\end{subfigure}
\begin{subfigure}[b]{0.5\textwidth}
\includegraphics[width=1.1\textwidth,trim={1cm 0 6 0},clip]{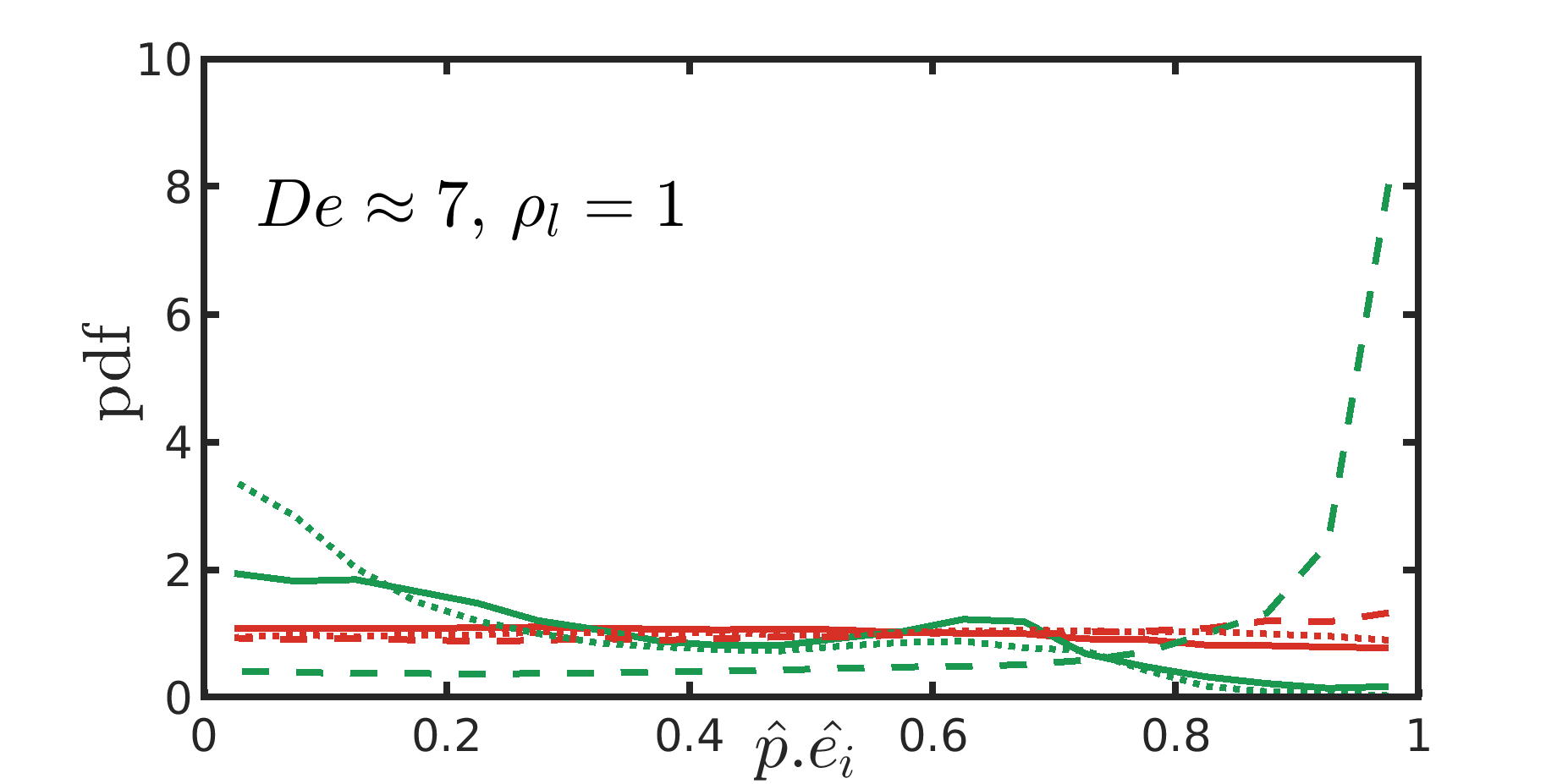}% Here is how to import EPS art
\caption{}
\end{subfigure}
\begin{subfigure}[b]{0.5\textwidth}
\includegraphics[width=1.1\textwidth,trim={1cm 0 6 0},clip]{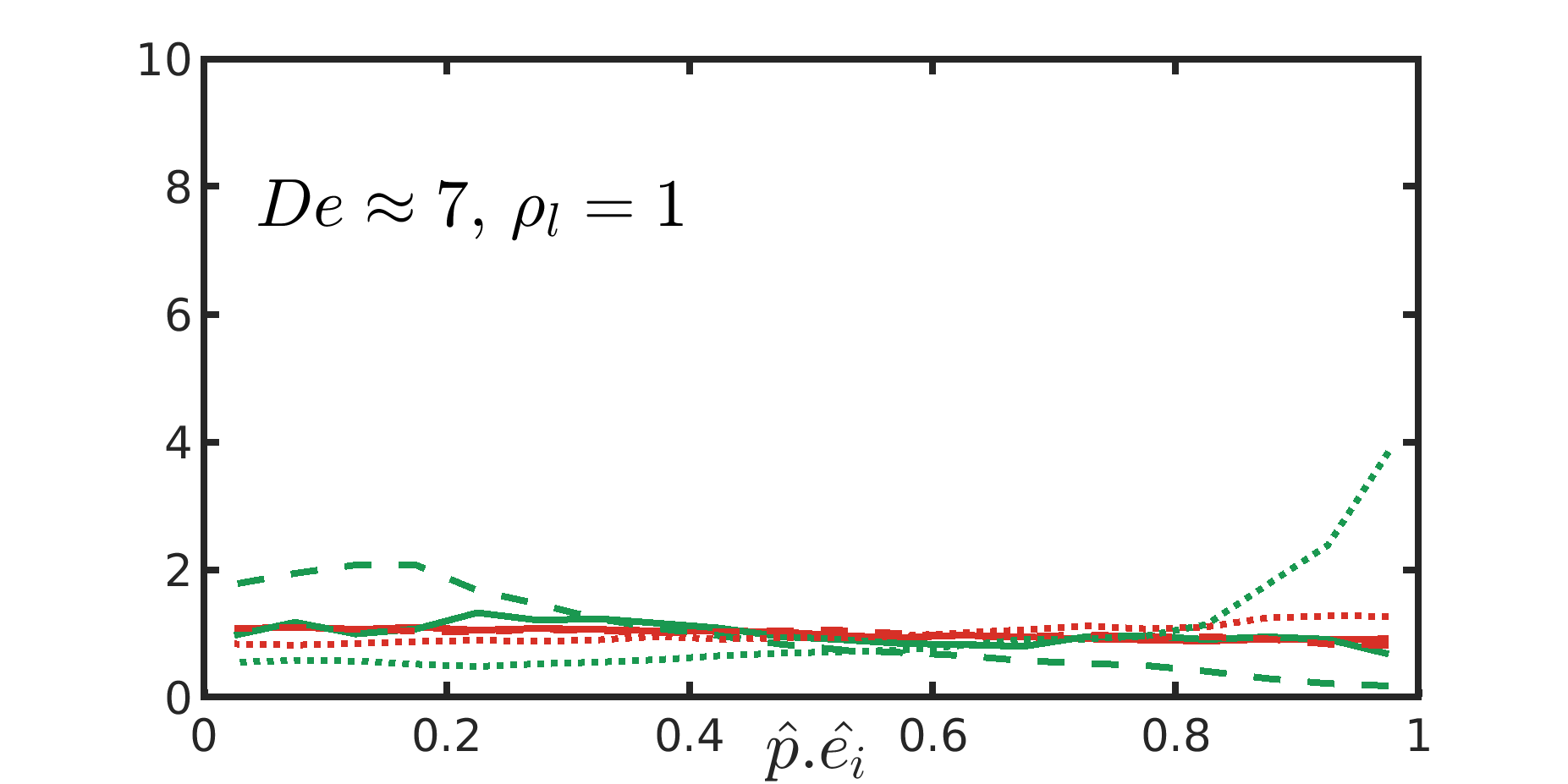}
\caption{}
\end{subfigure}
\begin{subfigure}[b]{0.8\textwidth}
\hspace{6cm}
\includegraphics[width=0.3\textwidth,trim={0cm 0 0 0},clip]{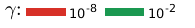}% Here is how to import EPS art
\end{subfigure}
\caption{Probability density functions of the alignment of denser-than-fluid fibers with the principal directions of (left) the strain rate tensor and (right) the conformation tensor, at $De \approx$ (a,b) 1, (c,d) 3, and (e, f) 7. Solid, dashed and dotted lines correspond to $i_{min}$, $i_{inter}$, and $i_{max}$, respectively.}
\label{aligndense}
\end{figure}
We now try to understand if there is a preferential alignment of the fiber towards specific flow quantities such as the principal directions of strain rate or that of the conformation tensor. \textcolor{black}{In turbulent mixing, it is of interest usually to understand how material fluid elements orient with components of velocity gradient tensor \citep{guala2006stretching}.} Flow-induced fiber alignment has a significant influence on the properties of composite materials, such as those made by processes like injection molding. \textcolor{black}{This property is also highly dependent on the configuration, eg: in HIT flows, fibers are known to usually align with intermediate eigenvector of strain rate tensor whereas for channels, they are a function of the channel height and is largely influenced by the coherent structures \citep{cui2021alignment}. Clearly, the alignment can also be influenced by viscoelastic stresses;} experiments with low Reynolds number simple shear flows have shown that orientation of semi-dilute fiber suspensions in weakly \citep{iso1996orientation1} and highly \citep{iso1996orientation2} elastic fluids  change in comparison to their aligned directions in Newtonian flows \citep{stover1992observations}. Further, this trend itself showed a complex set of behaviors based on  the fiber concentration \citep{iso1996orientation1, iso1996orientation2}. 

The fiber's local orientation (considering segments connecting two Lagrangian points) with the local fluid flow is computed instantaneously, and after adopting a coarse-graining procedure, the existence of a preferential alignment with any of the principal directions ($\hat{e}_{i}$) of the strain rate  and of the conformation tensor are explored. The three principal directions are chosen as $i_{min}, i_{inter}, i_{max}$ corresponding to eigenvalues $\chi$ such that $\chi_{min}<\chi_{inter}<\chi_{max}$, respectively. Usually, the eigendirection corresponding to $\chi_{max}$ corresponds to the most extensional direction and $\chi_{min}$ corresponds to the eigenvalue of the least extensional direction.

In Fig \ref{aligniso}, we consider only the neutrally-bouyant fibers and show the alignment of the fiber's orientation with the strain rate principal directions (left panel) and with the conformation tensor principal directions (right panel) at $De$ $\approx$ 0.3, 1, 7 (top to bottom). The solid, dashed, and dotted lines correspond to $i_{min}, i_{inter}, i_{max}$, respectively. Two extreme values of $\gamma$ are chosen, corresponding to a flexible ($\gamma = 10^{-8}$, red color) and rigid ($\gamma = 10^{-2}$, green color) fiber. It can be seen that for all three Deborah numbers, the fibers are mainly aligned with the intermediate eigenvectors ($i_{inter}$) of the strain rate and with the most extensional direction ($i_{max}$) of the conformation tensor. The anti-alignment with the least extensional direction ($i_{min}$) persists instead in both cases. It is also interesting to note that the variations of these trends for flexible and rigid fibers are negligible. Statistics of alignment of neutrally buoyant slender, microscopic fibers in homogeneous isotropic Newtonian flows \citep{pumir2011orientation,ni2015measurements} %(at lower Relambdas compared to this work=83,170) 
showed a tendency to align with ${i}_{inter}$, %and ew)
to be mostly perpendicular to ${i}_{min}$, and
no particular alignment with ${i}_{max}$, in agreement to the left panel here. Despite that, in polymeric flows the fibers are seen to align more with the most extensional direction of the polymeric tensor.

Major differences observed in the alignment of the denser-than-fluid fibers (Fig \ref{aligndense}) are that (i) the flexible and rigid fibers behave differently and that (ii) the alignment of the fibers with the polymeric tensor principal directions changes with the Deborah number, although with lower probabilities compared to their alignment with the velocity gradient tensor. At $De \approx 0.3$ (Fig \ref{aligndense}a), the fiber is still aligned with the intermediate eigenvector of the strain tensor, but it is anti-aligned with the most and least extensional direction. This alignment and anti-alignment trend is modified with the conformation tensor (Fig \ref{aligndense}b), where the fiber is anti-aligned with the intermediate direction and aligned with the least and most extended directions. This behaviour  is visible only for the most rigid fibers, while the flexible ones are nearly uniformly distributed, indicating no preferential direction. %In both cases, these effects become more pronounced as $\gamma$ increases, to the extent that the fibers become aligned with the least stretched direction for high $\gamma$s, from the most stretched direction at low $\gamma$s.
Further, as $De$ increases, the trends with respect to the strain rate matrix hold similar to the previous case (Fig \ref{aligndense} c, e), while with the conformation tensor we see a change: the alignment of the rigid fibers with the most extended direction vanishes at $De \approx 1$ (Fig \ref{aligndense}d), and returns at the larger value of $De$ (Fig \ref{aligndense} f). In short, the preferential alignment of the denser-than-fluid fibers is a function of the rigidity as well as the Deborah number. Also, consistently with the previous observations of this study, they do not quite follow the polymeric extensional direction as the neutrally-bouyant fibers.

\textcolor{black}{\cite{dotto2019orientation} investigated the orientation of inertial flexible fibers in channel flow, for different length and Stokes number of the fiber. %Fibers can bend and twist under the action of local fluid velocity gradients. The bending of flexible fibers increase with inertia.
Although a function of the channel height, it was reported that the fiber mean orientations in the streamwise and wall-normal directions showed a lesser tendency to preferentially align with the flow as the fiber inertia increased. The authors also mention that the fiber flexibility only plays a secondary role when fiber inertia is large enough. Analogously, here we also observe that an increase in fiber inertia influences the alignment of fibers and that flexible fibers show no preferential alignment.} %(is the bending of inertial fibres contradicting with our osbrvns?).
\subsubsection{Preferential sampling}
\begin{figure}
 \begin{subfigure}[b]{1\textwidth}
\centering \includegraphics[width=1\textwidth,trim={3cm 0 55 0},clip]{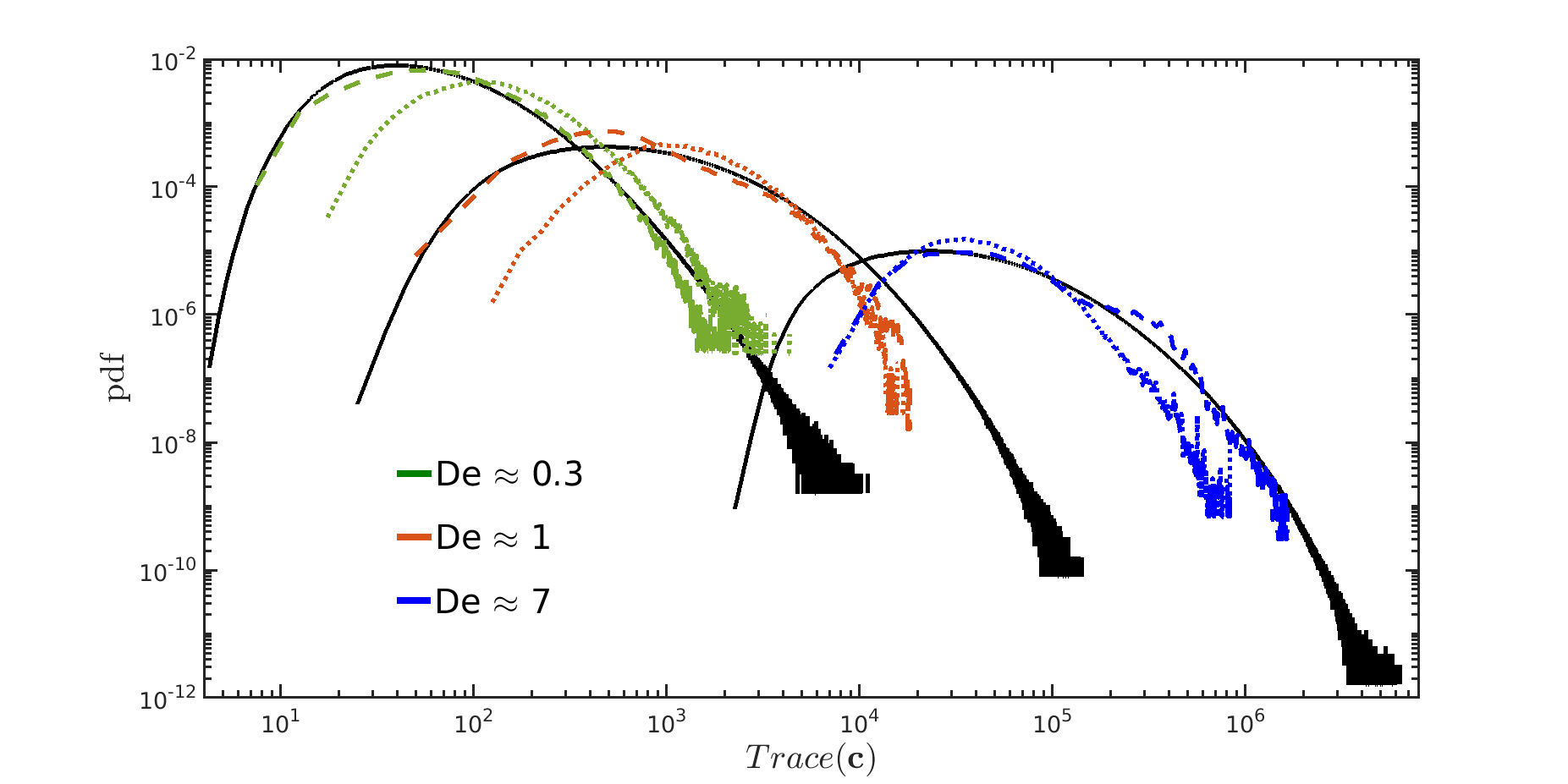}% Here is how to import EPS art
\caption{}
\end{subfigure}
 \begin{subfigure}[b]{0.5\textwidth}
\centering \includegraphics[width=1\textwidth,trim={1cm 0 55 0},clip]{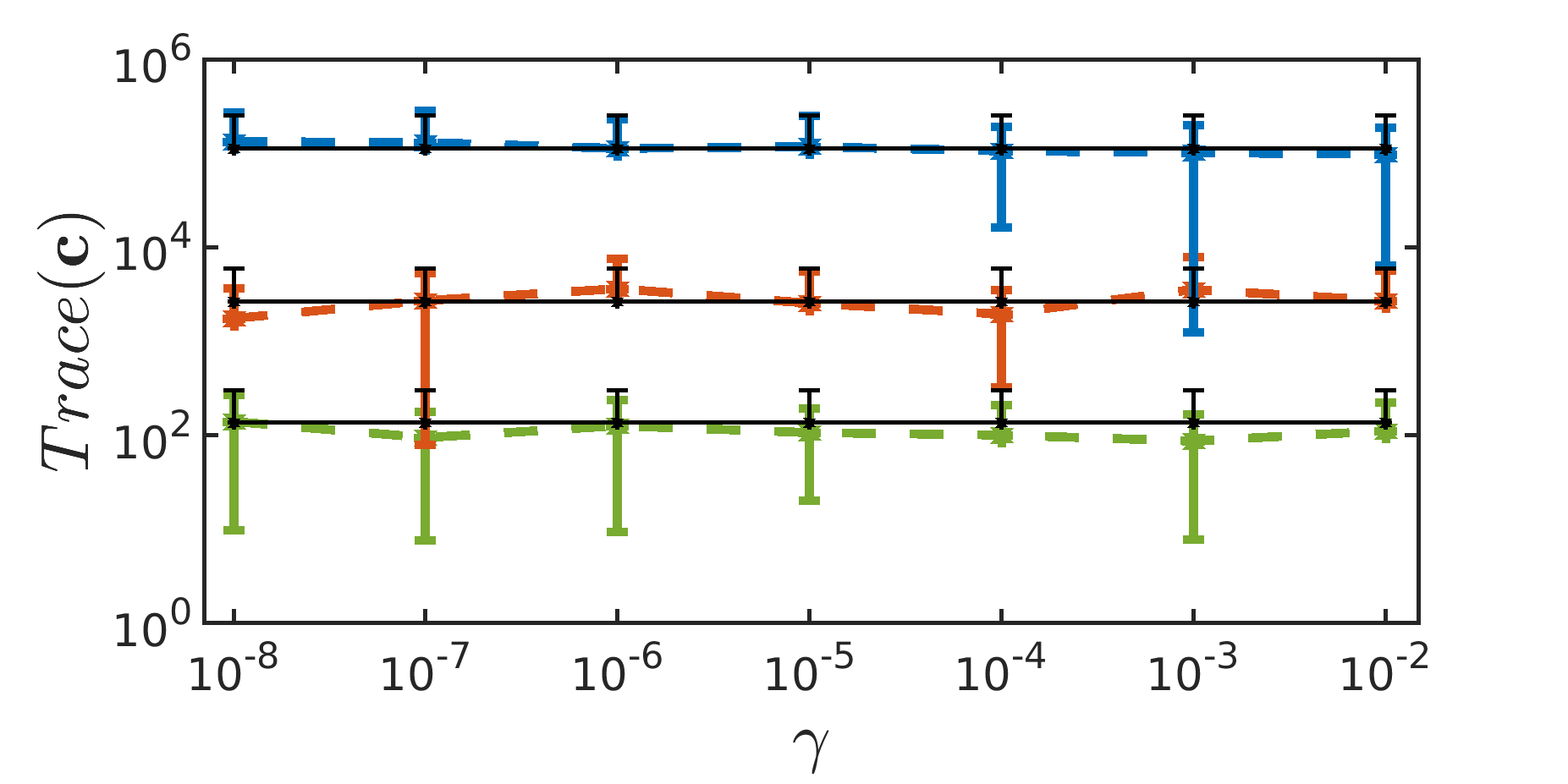}% Here is how to import EPS art
\caption{}
\end{subfigure}
%\quad
\hspace{0 cm}
 \begin{subfigure}[b]{0.5\textwidth}
\centering \includegraphics[width=1\textwidth,trim={1cm 0 45 0},clip]{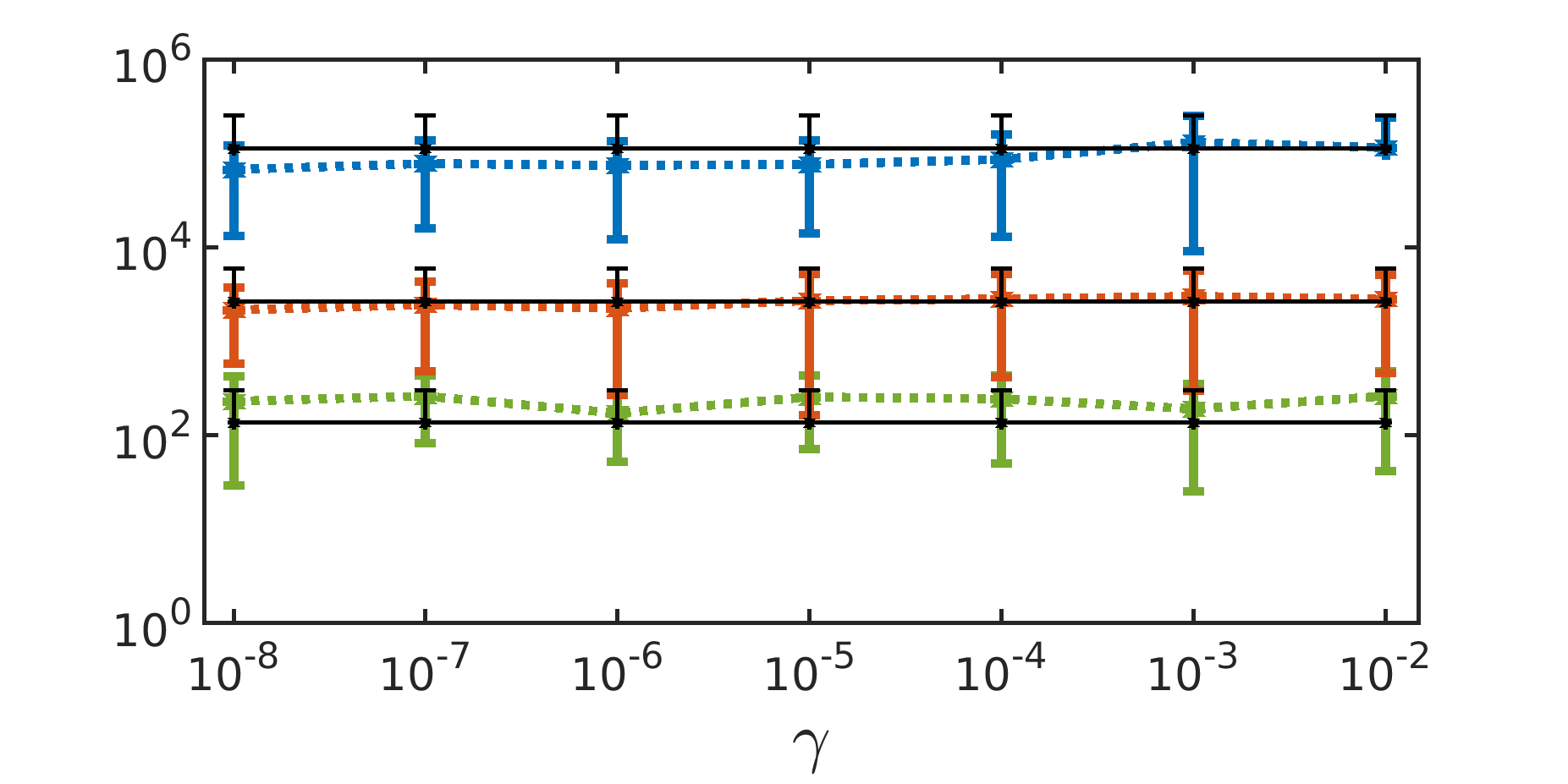}% Here is how to import EPS art
\caption{}
\end{subfigure}
\caption{(a) Probability density function of the trace of the conformation tensor (black) measured in the whole domain and (colors) perceived by the fibers, i.e. measured in a Lagrangian way. Dashed and dotted lines correspond to the neutrally-bouyant and denser fibers, both at $\gamma = 10^{-8}$. (b, c) Mean and standard deviation of the probability density function of the trace of the conformation tensor for (left) neutrally-bouyant and (right) denser fibers. The black lines represent the mean values of the Eulerian data.}
\label{trace}
\end{figure}
Finally, we investigate the preferential sampling of fibers. To do so, we compare the trace of the conformation tensor around the fibers (measured in a Lagrangian way) and the one in the whole domain (measured in an Eulerian way). Note that, the trace is a direct quantifier of the extension of the polymers, and the objective here is to find out to what extent the fibers see the polymer stretching. Fig \ref{trace}a shows the trace of the conformation tensor experienced by the fibers for the three different values of $De$, at a fixed value of $\gamma = 10^{-8}$. The black curves correspond instead to the Eulerian data. Both shift towards the right to higher values with increasing Deborah number, a direct effect of the increasing polymer extension with $De$. The probability distribution functions of the trace of the conformation tensor in the domain exhibit a very wide distribution at all $De$, positively skewed. The Lagrangian distributions show a similar shape, but with larger mean values and reduced tails; thus, fibers do not sample the whole regions of maximum polymer stretching. Denser fibers tend to feel more regions of high polymer extension at low $De$, while the opposite is evident at larger Deborah numbers. Fig \ref{trace}b, c  shows the mean and standard deviation of the neutrally-bouyant and denser fibers, respectively. Evidently, there is not much effect on the stiffness of the fiber, whereas the rate of increase in trace with respect to $De$ is slightly higher for the neutrally-bouyant fibers. Nevertheless, the fibers not sampling the regions with extreme polymer stretching can be the reason for the moderate effect of the Deborah number on their dynamics. 
%\begin{figure}[H]
  %\centerline{\includegraphics{Fig1}}% Images in 100% size
 %\caption{Trapped-mode wavenumbers, $kd$, plotted against $a/d$ for
    %three ellipses:\protect\\
    %---$\!$---,
%    $b/a=1$; $\cdots$\,$\cdots$, $b/a=1.5$.
%\label{fig:ka}
%\end{figure}

%\subsection{Figures}

%\begin{figure}
  %\centerline{\includegraphics{Fig2}}
  %\caption{The features of the four possible modes corresponding to
 % (\textit{a}) periodic\protect\\ and (\textit{b}) half-periodic solutions.}
%\label{fig:kd}
%\end{figure}

%\subsection{Tables}

%\begin{equation}
%\left. \begin{array}{ll}
%\displaystyle\frac{\p G_s}{\p y}=0
 % \quad \mbox{on\ }\quad y=0,\\[8pt]
%\displaystyle  G_a=0
%  \quad \mbox{on\ }\quad y=0,
% \end{array}\right\}
%  \label{symbc}
%\end{equation}

%Note that when equations are included in definitions, it may be suitable to render them in line, rather than in the equation environment: $\boldsymbol{n}_q=(-y^{\prime}(\theta),
%x^{\prime}(\theta))/w(\theta)$.
%Now $G_a=\squart Y_0(kr)+\Gat$ where
%$r=\{[x(\theta)-x(\psi)]^2 + [y(\theta)-y(\psi)]^2\}^{1/2}$ and $\Gat$ is
%regular as $kr\ttz$. However, any fractions displayed like this, other than $\thalf$ or $\squart$, must be written on the line, and not stacked (ie 1/3).
%wvt

\section{Conclusions}

We perform direct numerical simulations to explore the dynamical properties of fibers (both flexible and rigid, neutrally-bouyant and denser-than-fluid fibers) dispersed in a viscoelastic turbulent flow where fluid inertia and polymer elasticity are simultaneously present, with the three phases -- fluid, polymers and fibers -- which are fully coupled. The Reynolds number is sufficiently high to show a clear inertial range of scale which is altered when the Deborah number is increased. The goal is to examine how the microscopic polymers can influence the dynamics of macroscopic fibers, and to identify if the fibers reflect the changes in the fluid due to the polymers. For this, the flapping frequency of the fibers, their curvature, preferential alignment with and sampling of the flow were tracked for various values of the Deborah number.

We observe that the fibers primarily flap with a variety of time-scales, transpiring from the flow or their structural natural response, depending on the particle inertia, stiffness, and the polymer relaxation time. An examination of the frequency spectrum of the end-end displacement of the fibers showed that the expected time scale due to polymer stretching is not explicitly picked up by the fiber. Still, the neutrally bouyant fibers are weakly reflective of the polymer effects as the dominant flapping time scales of the fiber are also seen to increase with the polymer relaxation time; on the other hand, the denser-than-fluid fibers always oscillate with large time scales when flexible and with their natural frequency when sufficiently rigid, irrespective of the changes in the Deborah number. Polymer relaxation time impacts the neutrally-bouyant fiber curvature quantitatively but does not significantly impact their transition to buckling; on the contrary, the denser fibers -- which also exhibit larger curvatures -- are passive to the variations of the relaxation time. The neutrally-bouyant fibers show a high level of alignment with the polymer conformation tensor, unresponsive to variations in their rigidity and the Deborah number. Conversely, the alignment of the denser fibers changes with both the Deborah number and the rigidity, especially with respect to the polymeric tensor, although these probabilities are much lower than the corresponding alignment with the strain rate tensor.

This study attempts for the first time to track the dynamical properties of long fibers fully coupled with viscoelastic high Reynolds number turbulent flows. It reveals a complex interplay between the fiber flexibility, the polymer relaxation time, and the fiber inertia in determining the response behaviors of the fibers. The study can be of interest to industries developing products/processes involving viscoelastic fluid-fiber suspensions to optimize manufacturing and quality control processes and is a fundamental addition to this field of study. \textcolor{black}{Future works should take into account the effect of gravity as well. \cite{ardekani2017sedimentation} studied the sedimentation of prolate spheroids in HIT with application to non-motile phytoplanktons, and showed that settling spheroids showed an increased mean settling speed from those in a queiscent fluid; the authors of this study suggest that flexural stiffness is a dynamically important attribute to diatom chains that should be taken into account in future studies. Also the ability of shear thinning fluid to modify particle sedimenting velocity has been demonstrated by \cite{alghalibi2020sedimentation} through DNS. Along similar lines, it would be compelling to track the observations made in the current work by considering sedimentation effects along with considering fiber stiffness and Deborah number effects. \cite{banaei2020inertial} found definite differences between rigid and flexible fibers with respect to their settling velocities and alignment with the direction of gravity, while \cite{rahmani2023stochastic} described the shapes of inertial settling flexible fibers of large aspect ratios. It would be an intriguing take to perceive these known dynamics of sedimenting fibers in the background of elasto-inertial turbulence.} 

\backsection[Acknowledgements]{The research was supported by the Okinawa Institute of Science and Technology Graduate University (OIST) with subsidy funding from the Cabinet Office, Government of Japan. The authors acknowledge the computer time provided by the Scientific Computing section of Research Support Division at OIST and the computational time provided by the High Performance Computing Infrastructure (HPCI) under the grants hp210025, hp220099.}

\backsection[Declaration of interests]{The authors report no conflict of interest.}

\backsection[Author ORCIDs]{\\
M. S. Aswathy, https://orcid.org/0000-0003-4586-7364 \\
M. E. Rosti, https://orcid.org/0000-0002-9004-2292}

\bibliographystyle{jfm}
% @article{zhang2000flexible,
%   title={Flexible filaments in a flowing soap film as a model for one-dimensional flags in a two-dimensional wind},
%   author={Zhang, Jun and Childress, Stephen and Libchaber, Albert and Shelley, Michael},
%   journal={Nature},
%   volume={408},
%   number={6814},
%   pages={835--839},
%   year={2000},
%   publisher={Nature Publishing Group UK London}
% }
\bibliography{jfm}

\end{document}